\documentclass[12pt,preprint]{aastex}

\slugcomment{NAOJ-Th-Ap 2002, No.31; ApJ, in press}

\shorttitle{Nonthermal Emissions from Clusters of Galaxies}
\shortauthors{Fujita et al.}

\begin{document}

\title{Nonthermal Emissions from Particles Accelerated by Turbulence in
Clusters of Galaxies}

\author{Yutaka Fujita\altaffilmark{1,2},
Motokazu Takizawa\altaffilmark{3,4},
and
Craig L. Sarazin\altaffilmark{4}}

\altaffiltext{1}{National Astronomical Observatory, Osawa 2-21-1,
Mitaka, Tokyo 181-8588, Japan; yfujita@th.nao.ac.jp}
\altaffiltext{2}{Department of Astronomical Science, The Graduate
University for Advanced Studies, Osawa 2-21-1, Mitaka, Tokyo 181-8588,
Japan} 
\altaffiltext{3}{Department of Physics, Yamagata University,
Yamagata 990-8560, Japan} 
\altaffiltext{4}{Department of Astronomy,
University of Virginia, P.O. Box 3818, Charlottesville, VA 22903-0818}

\begin{abstract}
 We consider nonthermal emission from clusters of galaxies produced by
 particle acceleration by resonant scattering of Alfv\'{e}n waves driven
 by fluid turbulence through the Lighthill mechanism in the intracluster
 medium. We assume that the turbulence is driven by cluster mergers. We
 find that the resonant Alfv\'{e}n waves can accelerate electrons up to
 $\gamma\sim 10^5$ through resonant scattering. We also find that the
 turbulent resonant acceleration can give enough energy to electrons to
 produce the observed diffuse radio relic emission from clusters if the
 clusters have a pool of electrons with $\gamma\sim 10^3$. 
 This mechanism can also explain the observed hard X-ray emission from
 clusters if the magnetic field in a cluster is small enough ($\lesssim
 \mu G$) or the fluid turbulence spectrum is flatter than the Kolmogorov
 law. The fluid turbulence could be observed with {\it Astro-E2} in the
 regions where diffuse radio emission is observed. Although
 non-gravitational heating before cluster formation (preheating)
 steepens a relation between radio luminosity and X-ray temperature, our
 predicted relation is still flatter than the observed one.
\end{abstract}


\section{Introduction}
\label{sec:intro}

The intracluster medium (ICM), the largest detected baryonic component
of clusters of galaxies, contains thermal gas characterized by
temperatures in the range of $\sim 2-10$~keV and by central densities of
$\sim 10^{-3}\rm\; cm^{-3}$.  In addition to the thermal plasma,
observations have showed that the ICM contains nonthermal elements.
Diffuse synchrotron emission from the ICM has been observed in many
clusters \citep[e.g.,][]{kim90,gio93,gio00,kem01}.  Diffuse radio
sources in clusters are often classified either as peripheral cluster
radio relic sources, or central cluster radio halo sources.  These radio
halo and relic sources are not associated directly with individual
galaxies, but appear to be produced by relativistic electrons in the
intracluster space.  The typical energy of the electrons that produce
the radio emission depends on the observed frequency and on the
intracluster magnetic field, but it is estimated to be several
GeV. Moreover, hard X-ray emission has been detected as a nonthermal
tail at energies $\ga$20 keV in at least two clusters. The Coma cluster
was detected with both {\it BeppoSAX} \citep{fus99} and the {\it Rossi
X-Ray Timing Explorer} ({\it RXTE} ; \citealp*{rep99}). Possible weaker
excesses may have been seen in Abell~2199 \citep{kaa99}, Abell~2319
\citep{gb02}, and Abell~3667 \citep{fus01}. This radiation is often
believed to be inverse Compton emission, in this case produced by
relativistic electrons with the energy of $\sim 5$~GeV and Lorentz
factors $\gamma \sim 10^4$.

There are several possible origins for the relativistic electrons
\citep{ato00,ens01,ens02}. Electron acceleration at shocks in the ICM
has been the most popular idea
\citep[e.g.,][]{col98,byk00,tot00,wax00,ost02}. These shocks may be
attributed to the interaction between jets originated from active
galactic nuclei (AGNs) and ICM, to galactic winds, or to the accretion
of intergalactic gas.  Since most of the observed radio sources are very
large and are often found in irregular clusters, they may be attributed
to shocks formed by cluster mergers
\citep*{roe99,bla01,buo01,fuj01d}. \citet{tak00} calculated the
nonthermal emission from relativistic electrons accelerated around the
shocks produced during a merger of clusters with equal mass. They found
that the hard X-ray and radio emission is luminous only while signatures
of merging events are clearly seen in the ICM. For some clusters,
observations suggest that electrons are actually accelerated by
shocks. For example, \citet{mar01a} showed that a high-temperature
shocked region is associated with diffuse radio emission in Abell~665 by
analyzing data from the {\it Chandra} Observatory.  Shock acceleration
or re-acceleration seems one of the most promising candidates of the
source of relativistic particles in the localized, peripheral radio
relic sources.

On the other hand, some clusters have large diffuse radio emission that
does not seem to be associated with shocks.  Since the lifetimes of the
particles which produce the radio emission in clusters are short, it is
unlikely that radio emission will be found far from shocks if particles
are only accelerated at shocks.  Since the large radio sources are
generally found in merging clusters, the simplest explanation would be
an acceleration mechanism associated with mergers which is not directly
related to shocks or which persists in the plasma for some period after
the passage of a merger shock.  For Abell~3667, a diffuse radio emission
is detected in the northwest of the cluster center
\citep{rot97}. Although this cluster has a rapidly moving substructure,
the Mach number is estimated to be only $\sim 1$ \citep*{vik01}. Thus,
even if there are shocks in the cluster, the efficiency of shock
acceleration may be too small to produce the observed synchrotron
spectrum within the standard theory of shock acceleration. Moreover,
although Abell~2256, which is known as a merging cluster, also has a
large diffuse radio emission \citep{rot94}, the X-ray temperature map
does not show shock features \citep{sun02}.  The viewing angle may
prevent us from observing the shock.  However, even if particles are
accelerated at a bow shock, the size of the radio emission ($\sim$Mpc)
seems to be too large for a cluster to cross within the lifetimes of the
particles responsible for the emission. Recently, \citet{gab02}
investigated the effect of the low Mach numbers of merger-related shocks
on particle acceleration at the shocks.  They suggest that major
mergers, which often invoked to be sites for the production of extended
radio emissions, have shocks which are too weak to result in appreciable
nonthermal activity.

Fluid turbulence in ICM is another possible origin of particle
acceleration \citep{bla00,bru01a,bru01b}. Numerical simulations show
that cluster mergers can generate strong turbulence in the ICM
\citep{roe99,ric01}. The fluid turbulence can induce MHD waves
\citep{lig52,kul55,kat68,eil84}, which accelerate particles through
wave-particle resonance \citep*{jaf77,rol81,sch87}. Contrary to shock
acceleration, the turbulent resonant acceleration does not require
strong shocks. Even if a subcluster is moving with a subsonic velocity
in a host cluster, the induced turbulence may produce high energy
particles and nonthermal emission. As the first step of this study, in
this paper we consider particle acceleration by Alfv\'{e}n waves, which
are easily generated by fluid turbulence and do not damp easily.  The
effects of other MHD waves may also be important
\citep{sch98,rag98}. Recently, \citet*{ohn02} studied the resonant
acceleration in the Coma cluster; we will study the turbulence and
acceleration in clusters more generally.  As will be shown in
\S\ref{sec:result}, turbulent resonant acceleration is not effective
when two clusters collide with a large relative velocity because the
lifetime of the induced fluid turbulence is shorter than the
acceleration time of the particles. Thus, turbulent acceleration is most
effective when two clusters approach from a relatively small initial
separation because the relative velocity when they pass each other is
then relatively small.  Moreover, this mechanism is more likely to apply
to peripheral relics rather than central halos because two clusters have
the largest relative velocity at the point of closest approach of the
cluster centers; this large velocity decreases the time-scale of the
fluid turbulence and also makes ram-pressure stripping of the gas of the
smaller cluster effective, which reduces the volume of the turbulent
region. We will concentrate on models and parameters appropriate for the
cases where this mechanism is likely to be most important.

This paper is organized as follows. In \S\ref{sec:model}, we summarize
our models. In \S\ref{sec:result} we give the results of our
calculations and compare them with observations. Conclusions are given
in \S\ref{sec:conc}.

\section{Models}
\label{sec:model}

\subsection{Fluid Turbulence, Lighthill Radiation, 
and Particle Acceleration}
\label{sec:turb}

We use the turbulent resonant acceleration model studied by
\citet{eil84}; we summarize it in this subsection. We assume that fluid
turbulence is induced by the motion of a smaller cluster in a larger
cluster and its energy spectrum is described by a power law,
\begin{equation}
\label{eq:Wf}
 W_f(\kappa) = W_f^0 \kappa^{-m} \:,
\end{equation}
where $\kappa=2\pi/l$ is the wavenumber corresponding to the scale $l$,
$W_f(\kappa) d \kappa$ is the energy per unit volume in turbulence with
wavenumbers between $\kappa$ and $\kappa + d \kappa$, and $W_f^0$ and
$m$ are the constants. If one expresses the turbulent spectrum in terms
of eddy size, the spectrum is represented by $W_f(l)\propto
l^{m-2}$. The cascade of the fluid turbulence extends from a largest
eddy size $l_0=2\pi/\kappa_0$ down to a smallest scale determined by
dissipation, $l_D\sim l_0 (Re)^{-3/4}$, where $Re$ is the Reynolds
number. Since most of the energy of fluid turbulence resides in the
largest scale, the total energy density of fluid turbulence is given by
$E_t\sim \rho v_t^2$, where $\rho$ is the fluid density and $v_t$ is the
turbulent velocity of the largest scale $l_0$. The normalization $W_f^0$
can be derived from the relation $E_t = \int W_f(\kappa)d\kappa$ and is
\begin{equation}
\label{eq:W0}
 W_f^0 = E_t \kappa_T^{m-1}/R \:, 
\end{equation}
where
\begin{equation}
R = \frac{1}{m-1}\frac{\kappa_0 W_f(\kappa_0)}{\kappa_T W_f(\kappa_T)}\;.
\end{equation}
In the above equations, $\kappa_T=2\pi/l_T$ and $l_T$ is the wavelength
below which Alfv\'{e}n waves are driven.

Fluid turbulence will generate Alfv\'{e}n waves via Lighthill radiation
\citep{lig52,kul55,kat68}. The maximum wavelength of Alfv\'{e}n waves
driven by the turbulence cannot be larger than $l_0$. Instead,
Alfv\'{e}n waves are expected to be driven at the wavelength at which
there is a transition from large-scale ordered turbulence to small-scale
disordered motions. Unfortunately, there is no absolute definition of
this transition.  Following \citet{eil84}, we adopt the Taylor length as
an estimate of the transition scale.  The Taylor length is given by
\begin{equation}
\label{eq:Taylor}
 l_T \equiv \left[\overline{v_{f,x}^2}/
\overline{\left(\frac{\partial v_{f,x}}{\partial x}\right)^2}
\, \right]^{1/2}
= l_0 (15/Re)^{1/2} \:,
\end{equation}
where $\bf{v}_f$ is the turbulent velocity vector, $x$ is the space
coordinate, and $v_{f,x}$ is the $x$-component of the vector $\bf{v}_f$
\citep{ten72,eil84}.  The Taylor scale is determined by the ratio of the
r.m.s.\ turbulent velocity ($[\overline{v_{f,x}^2} ]^{1/2}$), which is
determined by the large scale motions, to the r.m.s.\ velocity shear ($[
\overline{ \left( \partial v_{f,x}/\partial x \right)^2} ]^{1/2}$),
which is mainly due to small scale motions.

A fluid eddy of size $l$ has a velocity $v_f=|\bf{v}_f|$, 
\begin{equation}
\label{eq:vl}
 v_f(l)\approx \left[\frac{l W_f(l)}{\rho}\right]^{1/2}
= \left[\frac{E_t}{\rho R}
\left(\frac{\kappa}{\kappa_T}\right)^{1-m}\right]^{1/2} \:.
\end{equation}
Turbulence on a scale $\kappa$ will radiate Alfv\'{e}n waves at the
wavenumber
\begin{equation}
\label{eq:vfvA}
 k= [v_f(l)/v_A]\kappa \:,
\end{equation}
where $v_A=B/\sqrt{4\pi \rho}$ is the Alfv\'{e}n velocity in plasma with
the magnetic field $B$ and the gas density $\rho$. Let $v_f [ \kappa (
k ) ]$ be the fluid velocity on the fluid scale $\kappa ( k )$ which
drives Alfv\'{e}n waves of wavenumber $k$.  From equations~(\ref{eq:vl})
and~(\ref{eq:vfvA}), we obtain
\begin{equation}
\label{eq:vfk}
v_f [ \kappa (k) ] = v_A \left(\frac{E_t}{\rho v_A^2 R}\right)^{1/(3-m)}
\left(\frac{k}{\kappa_T}\right)^{(1-m)/(3-m)} \:.
\end{equation}

We assume that the energy going into Alfv\'{e}n waves at wavenumber $k$
is given by a power law
\begin{equation}
 I_A(k) = I_0 (k/\kappa_T)^{-s_t} \:,
\end{equation}
where $I_A (k) dk$ is the energy per unit volume per unit time going
into Alfv\'{e}n waves with wavenumbers in the range $k \rightarrow k +
dk$, and $I_0$ and $s_t$ are the constants. In this case, the total
power going into the Alfv\'{e}n mode from fluid turbulence is
\begin{equation}
\label{eq:IAint}
 P_A = \int_{k_T}^{k_{\rm max}} I_A(k)dk
\approx \frac{I_0 k_T}{s_t-1}\left(\frac{k_T}{\kappa_T}\right)^{-s_t} \:,
\end{equation}
where $k_T= [v_f(l_T)/v_A]\kappa_T$.  We have assumed that $k_T \ll
k_{\rm max}$ and $s_t>1$. On the other hand, according to the Lighthill
theory, $P_A$ is given by
\begin{equation}
\label{eq:IAori}
 P_A 
= \eta_A \frac{v_f(l_T)}{v_A} \times \rho v_f(l_T)^3 \kappa_T \:,
\end{equation}
where $\eta_A$ is an efficiency factor of order unity
\citep{kat68,hen82,eil84}. By comparing equations~(\ref{eq:IAint})
and~(\ref{eq:IAori}) and using equations~(\ref{eq:vfvA})
and~(\ref{eq:vfk}), it can be shown that
\begin{equation}
 s_t = \frac{3(m-1)}{3-m} \:,
\end{equation}
\begin{equation}
 I_0 = \eta_A (s_t-1) \rho v_A^3 
\left(\frac{E_t}{\rho v_A^2 R}\right)^{3/(3-m)} \:,
\end{equation}
and 
\begin{equation}
\label{eq:PA}
 P_A = \eta_A \left(\frac{E_t}{\rho v_A^2 R}\right)^2 
    \rho v_A^3 \kappa_T \:
\end{equation}
\citep{eil84}.

The Alfv\'{e}n waves respond to this energy input according to
\begin{equation}
\label{eq:WAevo}
 \frac{d W_A(k)}{dt} = I_A(k)- \gamma(k) W_A(k) \:,
\end{equation}
where $W_A(k) dk $ is the energy per unit volume in Alfv\'{e}n waves
with wavenumbers in the range $k \rightarrow k + dk$, and $\gamma (k)$
is the damping rate \citep{eil79,eil84}. We consider damping by the
acceleration of relativistic particles. For a particle with momentum
$p$, the resonance condition is given by
\begin{equation}
\label{eq:res}
 k \approx \Omega m_e/[p(\mu-v_A/c)] \:,
\end{equation}
where $\Omega = eB/m_e c$ is the nonrelativistic cyclotron frequency,
and $\mu=v_{\parallel}/c$ is the projected particle velocity along the
magnetic field normalized by the light velocity $c$ \citep{eil79}.
Then, the damping rate due to accelerating particles is
\begin{equation}
\label{eq:gamacc}
 \gamma_{\rm acc}(k) = -\frac{4 \pi^3 e^2 v_A^2}{c^2}\frac{1}{k}
\int_{p_{\rm min}(k)}^{p_{\rm max}} p^2 
\left[1-\left(\frac{v_A}{c}+\frac{\Omega m_e}{pk}\right)^2\right]
\frac{\partial f(p)}{\partial p}dp \:,
\end{equation}
where $e$ and $m_e$ are the electron charge and mass, and $p_{\rm
min}(k) = \Omega m_e/[k(1-v_A/c)]$ from the resonance condition
(eq.~[\ref{eq:res}]).  The electron phase space distribution function is
$f(p)$ \citep{eil79,eil84}. In general, $p_{\rm max} \gg p_{\rm min}$
(see equation~[\ref{eq:gmax}] below) and the electron phase space
distribution function declines rapidly at large $p$, so the upper limit
does not contribute significantly to the integral in
equation~(\ref{eq:gamacc}). In fact, although the integral converges
whenever $f(p)$ is steeper than $p^{-3}$, we later consider only the
case of $4\lesssim s \lesssim 5$, where $f(p) \propto p^{-s}$.

We consider the case where the wave and particle spectra are represented
by power-law functions, namely,
\begin{equation}
\label{eq:WApow}
 W_A(k) = W_A^0 k^{-\nu} \:,
\end{equation}
\begin{equation}
\label{eq:fpow}
 f(p) = f_0 p^{-s} \:,
\end{equation}
where $W_A^0$ and $f_0$ are the constants. Since the characteristic wave
response time is $\sim \gamma_{\rm acc}^{-1}$, which is short ($\sim
10^4$ yr) compared to other relevant time-scales in our model, we can
assume that $dW_A/dt=0$. Thus, from equations~(\ref{eq:WAevo})
and~(\ref{eq:gamacc}), and assuming $v_A \ll c$, we obtain
\begin{equation}
 W_A^0 = \frac{I_0}{f_0}\kappa_T^{s_t} \frac{s-2}{8 \pi^3} 
\frac{c^2}{e^2 v_A^2} (\Omega m_e)^{s-2} \:,
\end{equation}
\begin{equation}
 \nu = s+s_t-3 \;.
\end{equation}

On the other hand, the electron distribution evolves according to the
equation,
\begin{equation}
\label{eq:fevo}
 \frac{\partial f}{\partial t} = \frac{1}{p^2}
\frac{\partial}{\partial p}
\left[p^2 D(p) \frac{\partial f}{\partial p}+ S p^4 f\right]\:.
\end{equation}
Here, $S=4 (B^2 + B_{\rm CMB}^2) e^4/(9 m_e^4 c^6)$ is the inverse
Compton and synchrotron loss coefficient, and $B_{\rm
CMB}=3.25(1+z)^2\rm\; \mu G$.  In this paper, we assume that the
redshift of a model cluster is $z=0$. The diffusion coefficient owing to
the resonant scattering off random Alf\'ven waves is
\begin{equation}
 D(p) = \frac{2 \pi^2 e^2 v_A^2}{c^3}\int_{k_{\rm min}(p)}^{k_{\rm max}}
W_A(k) \frac{1}{k}
\left[1-\left(\frac{v_A}{c}+\frac{\Omega m_e}{pk}\right)^2\right]dk \:,
\end{equation}
where $k_{\rm min}(p)=\Omega m_e/[p(1-v_A/c)]$. The value of the upper
cutoff $k_{\rm max}$ is not important as long as $k_{\rm max}\gg k_{\rm
min}$ and $\nu>0$. Using equations~(\ref{eq:WApow}) and~(\ref{eq:fpow}),
equation~(\ref{eq:fevo}) is rewritten as
\begin{equation}
\label{eq:fevo2}
 \frac{\partial f}{\partial t} = \frac{1}{p^2} 
\frac{\partial}{\partial p}
\left(A p^{2+\nu}\frac{\partial f}{\partial p}+S p^4 f\right)
\end{equation}
with
\begin{equation}
 A = \frac{s-2}{2\pi\nu (\nu+2)}
\frac{(\Omega m_e)^{s-2-\nu}}{c}\frac{I_0}{f_0}\kappa_T^{s_t} \:.
\end{equation}
Equation~(\ref{eq:fevo2}) can further be rewritten as
\begin{equation}
 \frac{\partial f}{\partial t} = \frac{f(p)}{t_a(p)}
-\frac{f(p)}{t_e(p)}
\end{equation} 
with
\begin{equation}
\label{eq:ta}
 t_a(p) = [s(s-\nu-1)Ap^{\nu-2}]^{-1} \:,
\end{equation}
\begin{equation}
\label{eq:te}
 t_e(p) = [(s-4)Sp]^{-1} \;.
\end{equation}

If $\nu=3$, equations~(\ref{eq:ta}) and~(\ref{eq:te}) show that the
ratio
\begin{equation}
 \frac{t_a(p)}{t_e(p)} = \frac{1}{s}\frac{S}{A}
\end{equation}
is independent of $p$.  In other words, if and only if $\nu=3$, the form
of $f(p)$ does not evolve \citep{eil84}. Moreover, \citet{eil84} showed
that even if $\nu\neq 3$ initially, $W_A$ and $f$ evolve so that $\nu$
and $S/(sA)$ approach three and unity, respectively, on a timescale of
approximately $t_p \equiv \min[t_a, t_e]$. Note that \citet{eil84}
showed that equation (\ref{eq:fevo2}) has another time-independent
solution with $f(p)\propto p^{-4}$, in addition to the time-independent
solution with $f(p)\propto p^{-s}$ ($s=S/A$) which we adopt
here. However, the former solution requires externally supplied fluxes
of particles in contrast with the latter. Since we do not consider
external sources of particles in this paper, we ignore this other
solution.

 From equations~(\ref{eq:vfvA}) and~(\ref{eq:vfk}), we obtain
\begin{equation}
\label{eq:kTkT}
 k_T = \left(\frac{E_t}{\rho v_A^2 R}\right)^{1/2}\kappa_T
\end{equation}
Since the sum of synchrotron and inverse Compton emission rate is given
by
\begin{equation}
 P_{\rm SI} = \int_{p_1}^{p_{\rm max}} 4\pi S p^4 cf(p)dp \:,
\end{equation}
it can be shown that
\begin{equation}
 \frac{P_{\rm SI}}{P_A}=\frac{2s(s-2)}{15}
\left(\frac{k_T p_{\rm max}}{\Omega m_e}\right)^{5-s}
\frac{S}{sA}
\end{equation}
from equations~(\ref{eq:PA}) and~(\ref{eq:kTkT}) when $\nu=3$
\citep{eil84}. From the resonance condition (eq.~[\ref{eq:res}]),
\begin{equation}
\label{eq:pmax}
 p_{\rm max}\sim \Omega m_e/k_T \;.
\end{equation}
Moreover, as mentioned above, $S/(sA) \approx 1$ in a
steady-state. Thus, in this case,
\begin{equation}
\label{eq:PSI}
 \frac{P_{\rm SI}}{P_A}=\frac{2s(s-2)}{15}
\end{equation}
\citep{eil84}. We note that equations~(\ref{eq:IAori})
and~(\ref{eq:PSI}) indicate that the emission power, $P_{\rm SI}$,
mainly depends on the energy injection at the scale of $l_T$; it does
not much depend on the particle spectral index $s$, which is expected to
be of the order of 4. This means that even if the fluid spectrum index
$m$ for $\kappa>\kappa_T$ is different from that for $\kappa<\kappa_T$
because of the back reaction of particle acceleration \citep{ohn02}, the
luminosity of nonthermal emission does not change significantly,
although the emission spectrum may be affected.

\subsection{Viscosity of the ICM}
\label{sec:visc}

In order to apply the turbulence acceleration model in \S\ref{sec:turb},
we need to find the Reynolds number of the ICM, $Re= l_0 v_t/\nu_K$,
where $\nu_K$ is the kinetic viscosity. The viscosity in turn is
represented by $\nu_K=u_p \lambda_{\rm eff}/3$, where $u_p$ is the
thermal velocity of protons and $\lambda_{\rm eff}$ is the effective
mean free path of protons. For the transverse drift of protons in
magnetic fields, the mean free path is given by $\lambda_{\rm
eff}=\lambda_g^2/\lambda_c$ \citep*{spi62,ruz89}, where $\lambda_g$ is
the proton gyroradius and $\lambda_c$ is the mean free path due to
Coulomb collisions:
\begin{equation}
 \lambda_g = \frac{m_p u_p c}{e B}\;,
\end{equation} 
\begin{equation}
 \lambda_c = \frac{3^{3/2}(k_B T_{\rm gas})^2}
{4 \pi^{1/2} n_p e^2 \ln \Lambda}\;
\end{equation}
\citep{sar86}. In the above equations, $m_p$ is the proton mass, $k_B$
is the Boltzmann constant, $T_{\rm gas}$ is the gas temperature, $n_p$
is the proton density, and $\ln \Lambda$ is the Coulomb logarithm. Thus,
we have
\begin{equation}
 \nu_K = 1.3\times 10^5 {\rm\; cm^2\: s^{-1}}\:
\left(\frac{n_p}{\rm 10^{-3}\; cm^{-3}}\right)
\left(\frac{T_{\rm gas}}{\rm 10^8\; K}\right)^{-1/2}
\left(\frac{\ln \Lambda}{40}\right)
\left(\frac{B}{\rm 1\; \mu G}\right)^{-2}\:.
\end{equation}
We use this viscosity in \S\ref{sec:result}.

We would like to point out the relation between the viscosity and the
electron maximum energy.  As equation~(\ref{eq:pmax}) shows, the maximum
energy of accelerated electrons, $p_{\rm max}$, is related to
$l_T$. Thus, $p_{\rm max}$ should constrained by the viscosity $\nu_K$
or the Reynolds number $Re$ (eq.~[\ref{eq:Taylor}]). From
equations~(\ref{eq:vl}), (\ref{eq:vfvA}), and~(\ref{eq:pmax}), one can
show that
\begin{equation}
\label{eq:lT}
 l_T \approx \left[\frac{2\pi m_e c^2 l_0^{-(m-1)/2} 
v_t \gamma_{\rm max}}{B e v_A}\right]^{2/(3-m)} \;,
\end{equation}
where the maximum Lorentz factor is $\gamma_{\rm max}=p_{\rm max}/(m_e
c)$ for $p_{\rm max} \gg m_e c$.  Using equation~(\ref{eq:Taylor}) and
$Re=v_t l_0/\nu_K$, we obtain
\begin{eqnarray}
 \gamma_{\rm max} &=& \frac{e B^2 l_0}{4 c^2 m_e \pi^{3/2} v_t \rho^{1/2}}
\left(\frac{15\nu_K}{l_0 v_t}\right)^{(3-m)/4} \\
 &=& 1.2\times 10^5 \left(\frac{l_0}{300\rm\: kpc}\right)^{2/3}
\left(\frac{v_t}{200\rm\: km\: s^{-1}}\right)^{-4/3}
\left(\frac{B}{1\:\rm \mu G}\right)^2 \nonumber \\
 & & \times
\left(\frac{n_p}{10^{-3}\rm\: cm^{-3}}\right)^{-1/2}
\left(\frac{\nu_K}{1.3\times 10^5}\right)^{1/3} \\
 &=& 1.2\times 10^5 \left(\frac{l_0}{300\rm\: kpc}\right)^{2/3} 
\left(\frac{v_t}{200\rm\: km\: s^{-1}}\right)^{-4/3}
\left(\frac{B}{1\:\rm \mu G}\right)^{4/3} \nonumber \\
 & & \times
\left(\frac{n_p}{10^{-3}\rm\: cm^{-3}}\right)^{-1/6}
\left(\frac{T_{\rm gas}}{\rm 10^8\; K}\right)^{-1/6}
\left(\frac{\ln \Lambda}{40}\right)^{1/3} \, ,
\label{eq:gmax}
\end{eqnarray}
where the last two equations assume Kolmogorov turbulence ($m=5/3$).  If
nonthermal radio and hard X-ray emission from clusters is due to
turbulent acceleration, this upper-limit may explain the discrepancy
between the cluster magnetic fields derived from Faraday rotation
\citep[e.g.,][]{law82,kim90,gol93,cla01} and those derived from hard
X-ray emission assuming a power law particle spectrum without an upper
limit \citep{fus99,fus00}.  Previous work indicated that a high-energy
cutoff ($\gamma_{\rm max}\sim 10^4$) in the electron energy distribution
could resolve this discrepancy \citep{bru01a,pet01,fuj01d}. Our model
prediction (eq.~[\ref{eq:gmax}]) for the cutoff is close to that
required, although this is a very rough estimate.

\subsection{Cluster Merger}

We consider a merger in which a smaller cluster falls into a larger
cluster. The merger model is based on the galaxy infall model in
\citet{fuj01}.

\subsubsection{Mass Profile of a Cluster}

In this subsection, we do not discriminate between the smaller and
larger clusters unless otherwise mentioned.  The virial radius of a
cluster with the virial mass $M_{\rm vir}$ and the formation redshift
$z$ is defined as
\begin{equation}
 \label{eq:r_vir}
r_{\rm vir}=\left[\frac{3\: M_{\rm vir}}
{4\pi \Delta_c(z) \rho_{\rm crit}(z)}\right]^{1/3}\:,
\end{equation}
where $\rho_{\rm crit}(z)$ is the critical density of the universe and
$\Delta_c(z)$ is the ratio of the average density of the cluster to the
critical density at redshift $z$. The former is given by
\begin{equation}
\label{eq:rho_crit}
 \rho_{\rm crit}(z)
=\frac{\rho_{\rm crit,0}\Omega_0 (1+z)^3}{\Omega(z)}\:,
\end{equation} 
where $\rho_{\rm crit,0}$ is the critical density at $z=0$, and
$\Omega(z)$ is the cosmological density parameter.  For the Einstein-de
Sitter Universe, $\Delta_c(z)=18\:\pi^2$.

We assume that a cluster is spherically symmetric and ignore the gravity
of ICM. The density distribution of dark matter is represented by a
power-law,
\begin{equation}
 \label{eq:rho_m}
\rho_{\rm m}(r)=\rho_{\rm mv}(r/r_{\rm vir})^{-\alpha}\:,
\end{equation}
where $\rho_{\rm mv}$ and $\alpha$ are the constants, and $r$ is the
distance from the cluster center. The normalization, $\rho_{\rm mv}$, is
given by
\begin{equation}
\rho_{\rm mv}=\frac{3-\alpha}{3}\Delta_c\rho_{\rm crit}\:. 
\end{equation}
We choose $\alpha=2.4$, because the slope is consistent with
observations \citep*{hor99}. Moreover, the results of numerical
simulations show that the mass distribution in the outer region of
clusters is approximately given by equation (\ref{eq:rho_m}) with
$\alpha \sim 2.4$ \citep*{nav96,nav97}. We adopt a power-law rather than
the full NFW profile to avoid specifying a particular value for the
concentration parameter and its variation with cluster mass and
formation redshift.  At present, there is no complete and consistent
understanding of the variation of the concentration parameter based on
numerical simulations \citep{nav97,bul01}.  Moreover, the central cusp
that is a characteristic of the NFW profile does not have a significant
impact on the motion of a radially infalling object in the cluster
\citep[Fig.~2 in][]{fuj98}.  For example, if we assume that the cluster
mass within 0.5~Mpc from the center is $2.5\times 10^{14}\: M_{\sun}$
and that a smaller cluster falls into the center from an initial radius
of $r=2$~Mpc, the velocity difference at $r=0.1$~Mpc is less than 10\%
between an NFW profile with a scale radius of $0.9$~Mpc and our adopted
profile with $\alpha=2.4$.

We consider two ICM mass distributions. One follows equation
(\ref{eq:rho_m}) except for the normalization and the core structure;
\begin{equation}
\label{eq:ICM_G}
 \rho_{\rm ICM}(r)=\rho_{\rm ICM, vir}
\frac{[1+(r/r_{\rm c})^2]^{-\alpha/2}}
{[1+(r_{\rm vir}/r_{\rm c})^2]^{-\alpha/2}}\:.
\end{equation}
The ICM mass within the virial radius of a cluster is
\begin{equation}
\label{eq:M_ICM}
 M_{\rm ICM}=\int_{0}^{r_{\rm vir}}4 \pi r^2 \rho_{\rm ICM}(r)dr\:.
\end{equation}
The normalization $\rho_{\rm ICM, vir}$ is determined by the relation
$f_b =M_{\rm ICM}/M_{\rm vir}$, where $f_b$ is the gas or baryon
fraction of the universe. This distribution corresponds to the case
where the ICM is in pressure equilibrium with the gravity of the cluster
and is not heated by anything other than the gravity. We call this
distribution the `non-heated ICM distribution'. We introduce the core
structure to avoid the divergence of gas density at $r=0$ and assume
$r_c=0.1 r_{\rm vir}$. For the larger cluster into which a smaller
cluster falls, we will modify the core radius to take account of the
finite size of the smaller cluster (see \S\ref{sec:ram}). We use $f_b
=0.25 (h/0.5)^{-3/2}$, where the present value of the Hubble constant is
written as $H_0=100\:h\rm\: km\:s^{-1}\: Mpc^{-1}$. The value of $f_b$
is the observed ICM mass fraction of high-temperature clusters
\citep*{moh99,ett99,arn99}, for which the effect of non-gravitational
heating is expected to be small.

However, X-ray observations suggest that the ICM is also heated
non-gravitationally at least for nearby clusters
\citep*[e.g.,][]{dav96,pon99,llo00,xue00}.  Thus, we also model the
distribution of the heated ICM using the observed parameters of nearby
clusters as follows. In this paper, we assume that the ICM had been
heated before being accreted by clusters. However, the distribution will
qualitatively be the same even if the ICM is heated after accretion
\citep[see][]{loe00}.

Following \citet*{bal99a}, we define the adiabat
$K_0=P/\rho^{\gamma_{\rm ad}}$, where $P$ is the gas pressure, $\rho$ is
its density, and $\gamma_{\rm ad} =5/3$ is the constant adiabatic index.
If ICM had already been heated before accreted by a cluster, the entropy
prevents the gas from collapsing into the cluster with dark matter. In
this case, the ICM fraction of the cluster is given by
\begin{equation}
\label{eq:f_ICM}
f_{\rm ICM}=\min
\left[0.040\left(\frac{M_{\rm vir}}{10^{14}\:\rm M_{\sun}}\right)
\left(\frac{K_0}{K_{34}}\right)^{-3/2}
\left(\frac{t(z)}{10^9\rm\: yr}\right),
\:f_b \right]\:,
\end{equation}
where $K_{34}=10^{34}\rm\: erg\:g^{-5/3}\:cm^{2}$ \citep{bal99a}.

The virial temperature of a cluster is given by
\begin{equation}
 \frac{k_B T_{\rm vir}}{\mu m_p}
=\frac{1}{2}\frac{GM_{\rm vir}}{r_{\rm vir}}\:,
\end{equation}
where $\mu = 0.61$ is the mean molecular weight, and $G$ is the
gravitational constant. When the virial temperature of a cluster is much
larger than that of the gas accreted by the cluster, a shock forms near
the virial radius of the cluster \citep*{tak98,cav98}. The temperature
of the postshock gas ($T_2$) is related to that of the preshock gas
($T_1$) and is approximately given by
\begin{equation}
\label{eq:TT}
 T_2 = T_{\rm vir}+\frac{3}{2} T_1
\end{equation}
\citep{cav98}. We assume that the gas temperature does not change very
much for $r<r_{\rm vir}$ \citep[see][]{tak98}. In this case, the ICM
temperature of the cluster is given by $T_{\rm ICM}=T_2$. Since we
assume that the density profile of gravitational matter is given by
equation (\ref{eq:rho_m}) with $\alpha=2.4$, the density profile of ICM
is given by
\begin{equation}
\label{eq:ICM_H}
 \rho_{\rm ICM}(r)=\rho_{\rm ICM, vir}
\frac{[1+(r/r_{\rm c})^2]^{-3\beta/2}}
{[1+(r_{\rm vir}/r_{\rm c})^2]^{-3\beta/2}}\:, 
\end{equation}
where $\beta=(2.4/3)T_{\rm vir}/T_{\rm ICM}$
\citep[see][]{bah94}. Observations suggest that $T_1\sim 0.5-1$~keV
although it depends on the distribution of the gravitational matter in a
cluster \citep{cav98,fuj00}. We choose $3T_{1}/2=0.8$~keV hereafter.
The normalization $\rho_{\rm ICM, vir}$ is determined by the relation
$f_{\rm ICM}=M_{\rm ICM}/M_{\rm vir}$.

\subsubsection{Ram-Pressure Stripping}
\label{sec:ram}

We first consider a radially infalling smaller cluster accreted by a
larger cluster. Since we study the first infall of a smaller cluster
before relaxation, we will ignore the influence of the smaller cluster
on the gravitational potential of the larger cluster. (Nonthermal
emission from a relaxing cluster will be studied in \S\ref{sec:whole}.)
That is, we assume that the structure of the larger cluster remains
intact during the infall of the smaller cluster.  The exception is the
central region of the larger cluster. Since the smaller cluster has a
finite size, the gravitational effect of the central density peak in the
larger cluster will be smoothed out on the scale of the smaller
cluster's size or the tidal radius when the smaller cluster reaches the
center of the larger cluster. The tidal radius of the smaller cluster,
$\tilde{r}_{\rm ti}$, at the distance $r$ from the larger cluster center
is obtained by solving the equation,
\begin{equation}
 \frac{\tilde{r}_{\rm ti}}{r}
=\left[\frac{M_S(\tilde{r}_{\rm ti})}
{3 M_L(r)}\right]^{1/3} \;,
\end{equation}
where $M_S(r)$ and $M_L(r)$ are the mass profiles of the smaller and the
larger cluster, respectively \citep{bin87}. The internal velocity of the
smaller cluster is
\begin{equation}
 \tilde{v_{\rm ti}}=\sqrt{\frac{G M_S
(\tilde{r}_{\rm ti})}{\tilde{r}_{\rm ti}}} \:,
\end{equation}
and the dynamical time scale is $\tilde{t}_{\rm dyn}=\tilde{r}_{\rm
ti}/\tilde{v_{\rm ti}}$. As the smaller cluster approaches the larger
cluster center, $\tilde{r}_{\rm ti}$ decreases and the infall velocity
$V$ increases. However, when the infall time-scale $r/V$, is smaller
than $\tilde{t}_{\rm dyn}$, the tidal disruption virtually stops. We
define the minimum radius of the smaller cluster $\tilde{r}_{\rm min}$
as the tidal radius when $r/V=\tilde{t}_{\rm dyn}$ is satisfied
first. We give the core radius of the larger cluster by $r_{L,
c}=\max(0.1 r_{\rm vir},\tilde{r}_{\rm min})$ in
equations~(\ref{eq:ICM_G}) and~(\ref{eq:ICM_H}).
  
We investigate two cases for the initial position of the smaller
cluster. One is that the smaller cluster starts to move at the
turnaround radius of the larger cluster $r_{\rm ta}$. In this case, we
give the initial velocity of the smaller cluster, $v_i$, at the virial
radius of the larger cluster $r=r_{L, \rm vir}$, and it is
\begin{equation}
 \frac{v_i^2}{2}=\frac{G M_{L, \rm vir}}{r_{L, \rm vir}}
-\frac{G M_{L, \rm vir}}{r_{\rm ta}}\:,
\end{equation}
where $M_{L, \rm vir}$ is the virial mass of the larger cluster.
Assuming that $r_{\rm ta}=2 r_{L, \rm vir}$ on the basis of the virial
theorem, the initial velocity is
\begin{equation}
 v_i = \sqrt{\frac{G M_{L, \rm vir}}{r_{L, \rm vir}}}\:.
\end{equation}
The virial radius $r_{L, \rm vir}$ is given by equation (\ref{eq:r_vir}).

For the second case, we assume that the smaller cluster was initially
located near the center of the larger cluster, as expected for the
merger of substructure within a large cluster \citep{fuj02b}.  The
initial distance from the center of the larger cluster is given by $r_i
= x_i r_{L, \rm vir}$, where $x_i$ is a parameter and $0<x_i<1$; the
velocity at $r=r_i$ is zero.  For both cases, the velocity of the
smaller cluster is obtained by solving the equation of motion;
\begin{equation}
 \label{eq:motion}
\frac{dv}{dt}=-\frac{G M_L(r)}{r^2}\:.
\end{equation}
for $r>r_{L, c}$. We assume that $\rho_m(r)=\rho_m(r_{L,c})$ in
equation~(\ref{eq:motion}) for $r<r_{L, c}$.

As the velocity of a smaller cluster increases, the ram-pressure from
the ICM of the larger cluster affects the gas distribution of the
smaller cluster. \citet{sar02} indicated that the ram pressure affects
the gas distribution at radii which satisfy
\begin{equation}
\label{eq:ram}
 P_{\rm ram}\equiv \rho_{L, \rm ICM}(r)v^2 \gtrsim P_S(\tilde{r})\:,
\end{equation}
where $\rho_{L, \rm ICM}$ is the ICM density of the larger cluster,
$P_S$ is the static pressure of the smaller cluster, and $\tilde{r}$ is
the distance from the center of the smaller cluster. The static pressure
is given by $\rho_{S, \rm ICM}(\tilde{r}) k_B T_{S, \rm ICM}/(\mu m_p)$,
where the index $S$ indicates the smaller cluster. We assume that the
ICM of the smaller cluster is stripped outside the radius,
$\tilde{r}_{\rm st}$, satisfying $P_{\rm ram}=P_S(\tilde{r}_{\rm
st})$. However, if $\tilde{r}_{\rm ti}$ is smaller than $\tilde{r}_{\rm
st}$, we reset the ICM radius to $\tilde{r}_{\rm st}=\tilde{r}_{\rm
ti}$.

\subsubsection{Nonthermal Emission from Merging Clusters}

The motion of the smaller cluster relative to the ICM of the larger
cluster will induce turbulence behind the smaller cluster. The largest
eddy size and velocity are respectively $\lesssim \tilde{r}_{\rm st}$
and $\lesssim v$, where $v$ is the relative velocity of the smaller
cluster to the larger cluster. Thus, we define $l_0 = f_l\:
\tilde{r}_{\rm st}$ and $v_t = f_v v$, where $0<f_l<1$ and $0<f_v<1$. We
expect that nonthermal emission comes from the turbulent region
according to equation~(\ref{eq:PSI}). The volume of the turbulent region
is given approximately by
\begin{equation}
 V_{\rm turb} = \pi \tilde{r}_{\rm st}^2 v t_{\rm turb} \:,
\end{equation}
where $t_{\rm turb}=l_0/v_t$ is the time-scale of the turbulence. Thus,
the synchrotron luminosity and inverse Compton luminosity can be written
as
\begin{equation}
\label{eq:Lsyn}
 L_{\rm syn} = \frac{B^2}{B^2+B_{\rm CMB}^2} P_{\rm SI} V_{\rm turb}
\end{equation}
and
\begin{equation}
\label{eq:Lcom}
 L_{\rm IC} = \frac{B_{\rm CMB}^2}{B^2+B_{\rm CMB}^2} 
P_{\rm SI} V_{\rm turb} \;,
\end{equation}
respectively. We assume that large-scale magnetic fields are given by
\begin{equation}
\label{eq:mageq}
 \frac{B^2}{8 \pi}=b\times \frac{1}{2}\rho_{\rm ICM} v_t^2 \;,
\end{equation}
where $b$ is a parameter.

In our model, since the lifetimes of accelerated high-energy particles
are equal to or less than the time-scale of fluid turbulence (see
\S\ref{sec:gen}), the high-energy particles and the nonthermal emission
from them are mostly confined in the turbulent region with the volume of
$V_{\rm turb}$ behind a smaller cluster. Note that the local energy
density of the high energy particles is smaller than the thermal energy
density of ICM in the calculations in \S\ref{sec:result}.

\section{Results and Discussion}
\label{sec:result}

\subsection{General Features}
\label{sec:gen}

We calculate the nonthermal emission from a merging cluster using the
model presented in \S\ref{sec:model}. In this subsection, we study the
dependence of the nonthermal emission on the cluster masses.  The model
parameters are shown in Table~\ref{tab:para}. First, we study the case
where a smaller cluster falls from $r_i=0.5 r_{L, \rm vir}$, and we
adapt the non-heated ICM distribution (eq.~[\ref{eq:ICM_G}]) and
the Kolmogorov law ($m=5/3$) for the fluid turbulence (Model~A). From
now on, we assume $\Omega_0=1$ and $h=0.5$ for the sake of simplicity. A
larger cluster forms at $z=0$, and a smaller cluster forms at $z=0.5$.
We also assume that $f_l=0.5$ and $b=1$. \citet{kat68} showed that in
the case of equipartition of energy and eddy size between the velocity
and magnetic fields on small scales, $\eta_A=15.5/(2\pi)$. We adopt that
value for $\eta_A$.

\begin{figure}\epsscale{0.45}
\plotone{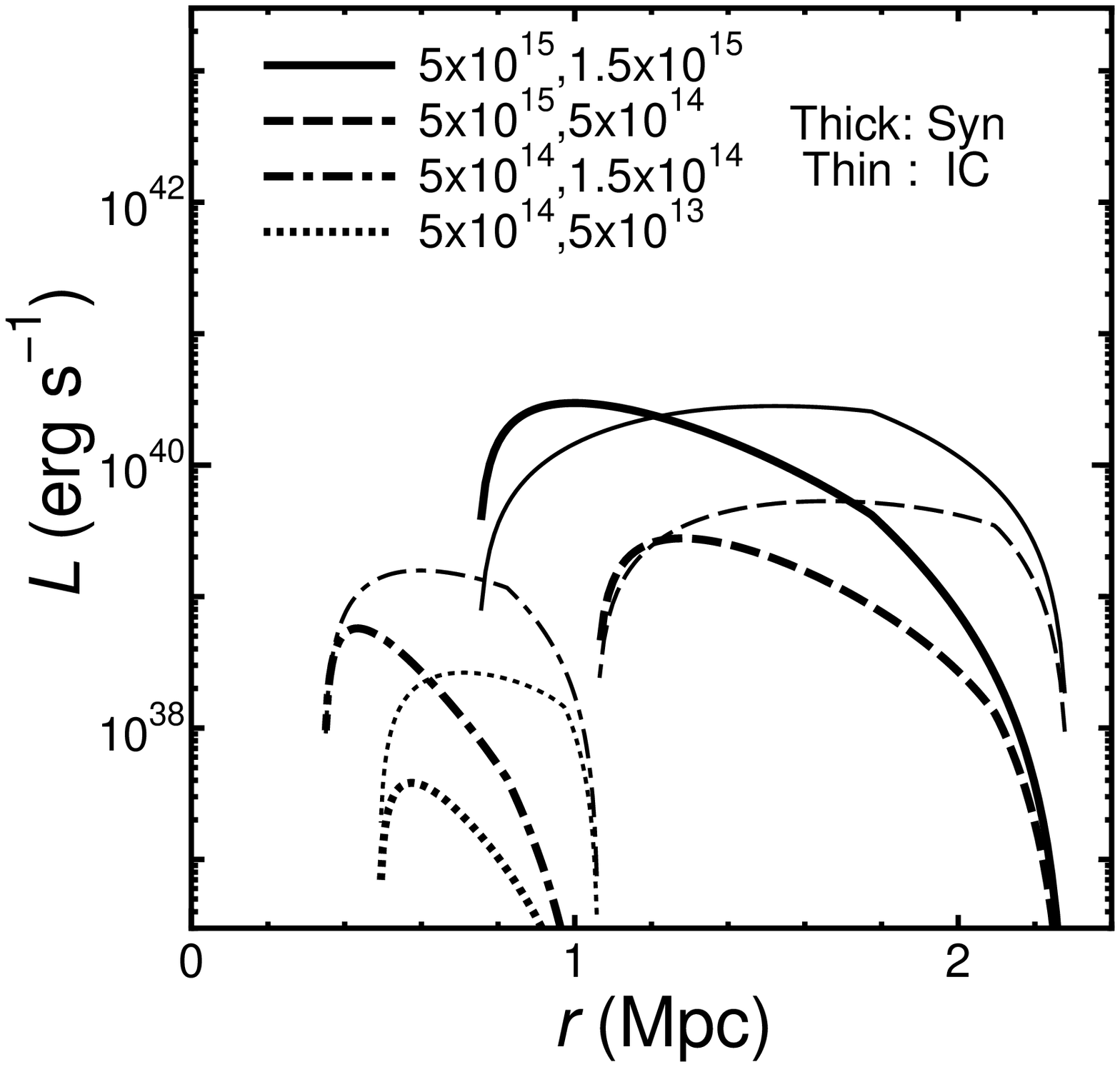} \caption{
 Synchrotron (thick lines) and
 inverse Compton (thin lines) luminosities from merging clusters when
 the smaller cluster is at the distance $r$ from the center of the
 larger cluster. The solid, dashed, dot-dashed, and dotted lines
 correspond to $(M_{L, \rm vir},M_{S, \rm vir})
 =(5\times 10^{15}, 1.5\times
 10^{15}), (5\times 10^{15}, 5\times 10^{14}), 
 (5\times 10^{14}, 1.5\times 10^{14}), (5\times 10^{14}, 5\times
 10^{13})\: M_{\sun}$, respectively.\label{fig:Asi}}
\end{figure}

Figure~\ref{fig:Asi} shows the synchrotron luminosity of merging
clusters when the smaller cluster is at the distance $r$ from the center
of the larger cluster. The outer bends in the curves (e.g. at
$r=1.8$~Mpc for the model of $M_{L, \rm vir}=5\times 10^{15} M_{\sun}$
and $M_{S, \rm vir}=1.5\times 10^{15} M_{\sun}$) correspond to the radii
at which the ICM radius of the smaller cluster equals to its tidal
radius. When $M_{L, \rm vir}=5\times 10^{15} M_{\sun}$ and $M_{S, \rm
vir}=1.5\times 10^{15} M_{\sun}$, the maximum synchrotron luminosity is
$\sim 3\times 10^{40}\rm\; erg\; s^{-1}$. Since the masses of the
clusters are close to the maximum observed so far, the synchrotron
luminosity should be the maximum of any clusters for the parameters we
have assumed.  This maximum luminosity is close to that obtained from
observations \citep[e.g.,][]{ens98,gio99,gio00,kem01}, which shows that
particle acceleration originating from fluid turbulence is a promising
candidate for the synchrotron radio emission from clusters.

The spectral index of the nonthermal emission is given by
$\alpha=(s-3)/2$. Since $s=4.5$ for $m=5/3$, the spectral index for the
model cluster is $\alpha=0.75$. This is somewhat smaller than the
observed values. For instance, for 18 clusters investigated by
\citet{kem01}, the average spectral index is $\alpha=1.2$ at
327~MHz. One possibility is that the upper cutoff of the energy
distribution of particles steepens the observed radio spectrum
(eq.~[\ref{eq:gmax}]). On the other hand, the observed steep spectrum
may be affected by the emission from the region where fluid turbulence
has almost disappeared and higher energy particles have lost their
energy. Although it is halo emission, \citet{gio93} showed that the
spectral index of the radio halo in the Coma cluster is flat at the halo
center ($\alpha\sim 0.5$) compared with that in the peripheral region
($\alpha\sim 1.5$). In the future, more observations of the spectral
index of halo and relic radio emission will be useful to study the
turbulent acceleration there.

Figure~\ref{fig:Asi} also shows the inverse Compton
luminosities. Compared to the synchrotron luminosities, they achieve
their maxima when the smaller cluster is in the outer region of the
larger cluster. This is because the magnetic fields in the clusters
increase with the kinetic energy of turbulence
(eq.~[\ref{eq:mageq}]), and the fraction of the inverse Compton
luminosity decreases as the magnetic fields increase
(eqs.~[\ref{eq:Lsyn}] and~[\ref{eq:Lcom}]). We present the magnetic
fields produced by a smaller cluster at the distance $r$ from the center
of a larger cluster in Figure~\ref{fig:Amag}. The absolute inverse
Compton luminosity ($\lesssim 3\times 10^{40}\rm erg\; s^{-1}$) we
predict in Figure~\ref{fig:Asi} is much smaller than the hard X-ray
luminosities observed in several clusters. For the Coma cluster, which
is known as a merging cluster, the {\it BeppoSAX} detected hard X-ray
flux of $2.2\times 10^{-11}$ ergs cm$^{-2}$ s$^{-1}$ in the $20-80$~keV
band.  This corresponds to the luminosity of $\sim 10^{43}$ ergs
s$^{-1}$ \citep{fus99}. For Abell~2256, which is also known as a merging
cluster, the flux of hard X-ray emission is $1.2 \times 10^{-11}$ ergs
cm$^{-2}$ s$^{-1}$ in the $20-80$~keV band range \citep{fus00} and the
corresponding luminosity is $\sim 10^{44}$ ergs s$^{-1}$. This may
suggest that the hard X-ray emission observed in clusters so far are not
attributed to the particles accelerated by turbulence, although our
turbulence model is favorable in the view of the maximum energy of
electrons (\S\ref{sec:visc}). In the future, it will be useful to
observe the spatial distributions of hard X-ray emission in a number of
clusters with sensitive detectors. If hard X-ray emission is observed
from the same region as synchrotron emission but there are no shock
features such as temperature jumps in this region, it is likely that the
emission is due to turbulent acceleration of electrons.

The fluid turbulence could be observed directly by detectors with high
spectral resolution. Figure~\ref{fig:Avt} shows the typical turbulent
velocity of the turbulence, $v_t$, induced by a smaller cluster at the
distance $r$ from the center of a larger cluster. The maximum velocity
is $\sim 300\rm\; km\;s^{-1}$ and it could marginally be detected by
observing broadened X-ray emission lines with {\it Astro-E2}; the
turbulence is likely to be observed at the regions where synchrotron
radio emission is observed.  Since it may be difficult to measure the
absolute line widths, it may be better to compare the line widths in
radio-emitting regions which are expected to be turbulent with those in
other regions of the cluster.  Moreover, although the typical turbulent
velocity is $\sim v_t$, there should be velocity components between
$v_t$ and the velocity of the smaller cluster, $v$. These components
would be observed as broad wings of the emission lines.  The spatial
scale of the turbulent nonthermal emission region ($v t_{\rm turb}$)
when the smaller cluster is at a radius $r$ is shown in
Figure~\ref{fig:Ascale}.

One concern is whether our assumption of steady turbulence and particle
acceleration can be justified or not. Figure~\ref{fig:Atturb} shows the
life span of the turbulence, $t_{\rm turb}\equiv l_0/v_t$. As can be
seen, $t_{\rm turb}\gtrsim 10^9$~yr for the outer regions of
clusters. On the other hand, we expect that the energy distribution of
accelerated electrons reaches the steady solution of $\nu = 3$ in the
time-scale of $\sim t_p = \min[t_a, t_e]$ (\S\ref{sec:turb}).  Since the
emission cooling time-scale, $t_e$, does not depend on an electron
energy spectrum, the relation $t_e < t_{\rm turb}$ is a sufficient
condition for the assumption of steady-state.  Figure~2 in \citet{sar99}
showed that the electrons with $\gamma\sim 300$ have the maximum
lifetime of several of $10^9$~yr, and that electrons with $\gamma\equiv
p/(m_e c)\gtrsim 10^3$ have the emission cooling times shorter than
$10^9$~yr. Thus, we think that the assumption of steady-state is
justified for electrons with $\gamma \gtrsim 10^3$ in the outer region
of a cluster. Thus, if the radio and hard X-ray nonthermal emission
from clusters is due to the electrons accelerated by turbulence up to
$\gamma\sim 10^4-10^5$, there must be a pool of source electrons with
$\gamma \sim 10^3$ to be accelerated by turbulence in clusters.  These
source electrons may be provided by the shock acceleration due to
cluster mergers and/or gas accretion onto clusters that occurred
$\lesssim 10^9$~yrs ago.  In particular, the merger shock wave expected
to form in front of the merging smaller cluster may provide those
electrons.  Alternatively, these electrons might come from AGNs in
clusters, or might be secondary electrons produced by the interactions
of cosmic-ray ions \citep{dol00,bla01,min01}.

As the smaller cluster reaches the center of the larger cluster, $t_{\rm
turb}$ decreases rapidly. In this region, since the fluid turbulence has
not developed fully to accelerate electrons with an energy of
$\gamma\sim 10^3$, the actual nonthermal luminosities of clusters may be
smaller than those in Figures~\ref{fig:Asi} even if there is a pool of
electrons with the energy of $\gamma\sim 10^3$. In other words, although
electrons with higher energy (say $\sim 10^4$) can be accelerated by
turbulence because of small $t_e$, the turbulent acceleration is not a
main contributor to the observed nonthermal emission.

Note that all the gas contained in the smaller cluster was stripped
before the smaller cluster reached the center of the larger cluster in
our model. Thus, there is no nonthermal emission from the central region
of the larger cluster (Figure~\ref{fig:Asi}).  This and the small
time-scale of the fluid turbulence in the central region suggest that
our model favors peripheral radio relics rather than central radio
halos. However, the ram-pressure stripping will depend on the central
structure of the smaller cluster, although the detailed study is beyond
the scope of this paper. If the smaller cluster has a distinct small
scale potential at the center as suggested by the observation of Fornax
cluster \citep{ike96}, the gas in it may not be removed easily. Also, if
the stars of the central galaxy in the small cluster can supply enough
gas, the gas may survive the ram pressure from the ICM of the larger
cluster. Moreover, the radiative cooling of the gas in the central
region of the smaller cluster may affects the ram-pressure stripping
\citep{rit02}. In fact, the cluster 1E~0657$-$56 has a rapidly moving
small component that appears to have passed the cluster center without
losing all of the gas \citep{mar02}.

\subsection{Parameter Dependence}
\label{sec:dep}

Since there are several uncertain parameters in our model, we
investigate the dependence of the nonthermal emission from
merging clusters on these uncertain parameters.

Figure~\ref{fig:Bsi} shows the nonthermal emission when $f_v=0.1$
and~0.5. The luminosities are small when $f_v$ is small. In particular,
the synchrotron luminosity varies more than the inverse Compton
luminosity. This is because the synchrotron luminosity is affected not
only by the energy injection from fluid turbulence but also by magnetic
fields. When $f_v$ or $v_t$ are small, the induced magnetic fields
(eq.~[\ref{eq:mageq}]) is also small. If $B<B_{\rm CMB}$, most of the
nonthermal emission should be the inverse Compton emission rather than
synchrotron emission (eqs.~[\ref{eq:Lsyn}] and~[\ref{eq:Lcom}]). On the
other hand, the results do not much depend on $f_l$.

\begin{figure}\epsscale{0.45}
\plotone{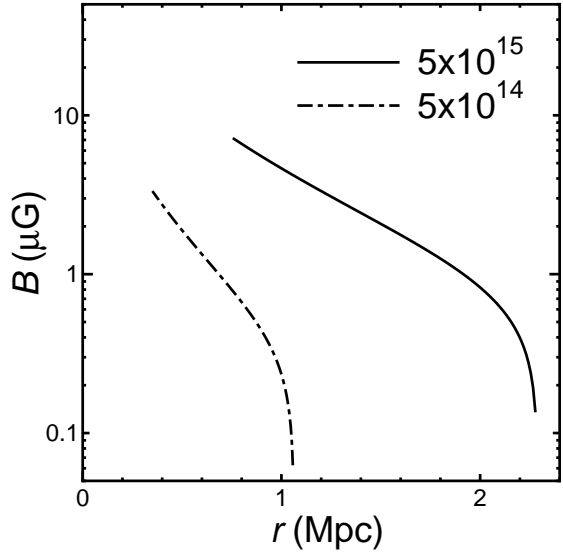} \caption{Magnetic fields generated by fluid
 turbulence when the smaller cluster is at the distance $r$ from the
 center of the larger cluster with the mass of 
 $M_{L, \rm vir}=5\times 10^{15}\:
 M_{\sun}$ (solid line) and $M_{L, \rm vir}=5\times 10^{14}\: M_{\sun}$
 (dot-dashed line).  \label{fig:Amag}}
\end{figure}

\begin{figure}\epsscale{0.45}
\plotone{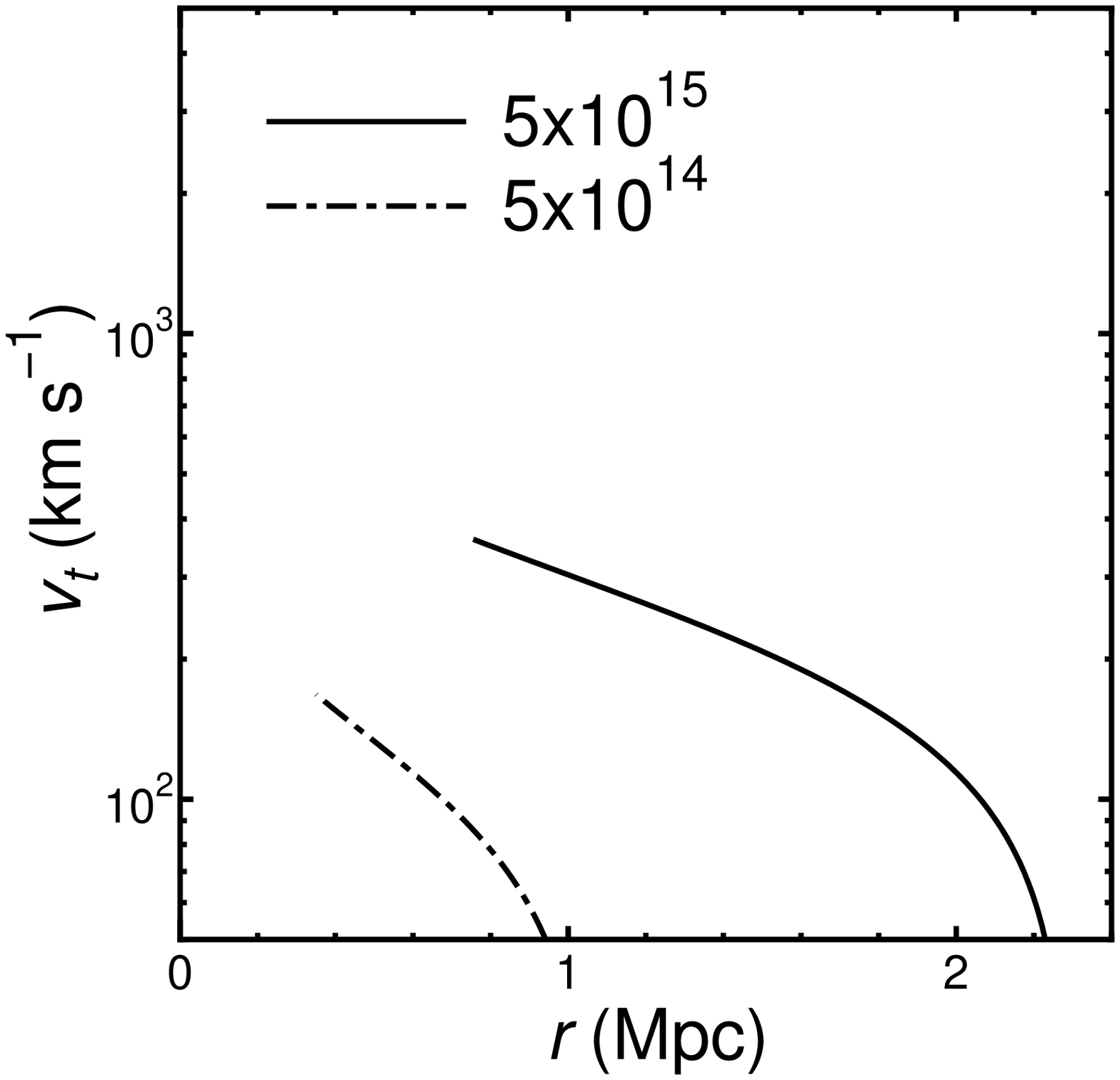} \caption{The velocity of fluid turbulence when the
 smaller cluster is at the distance $r$ from the center of the larger
 cluster with the mass of $M_{L, \rm vir}=5\times 10^{15}\: M_{\sun}$ 
 (solid
 line) and $M_{L, \rm vir}=5\times 10^{14}\: M_{\sun}$ (dot-dashed line).
 \label{fig:Avt}}
\end{figure}

\begin{figure}\epsscale{0.45}
\plotone{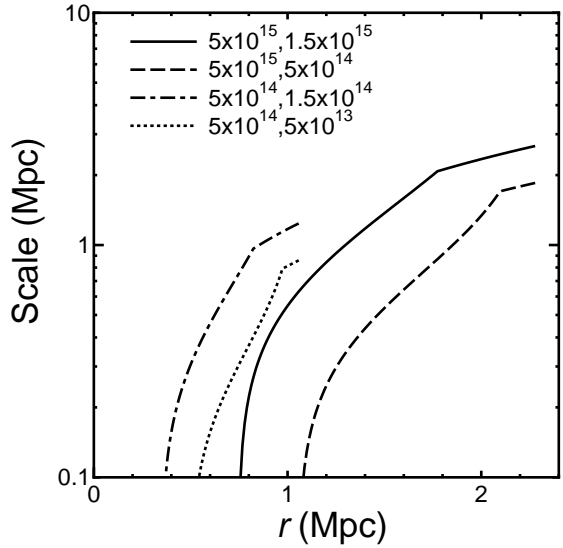} \caption{Spatial scale of nonthermal emission
 when the smaller cluster is at the distance $r$ from the center
 of the larger cluster.
 Notation is the same as Figure~\protect\ref{fig:Asi}.
\label{fig:Ascale}}
\end{figure}

\begin{figure}\epsscale{0.50}
\plotone{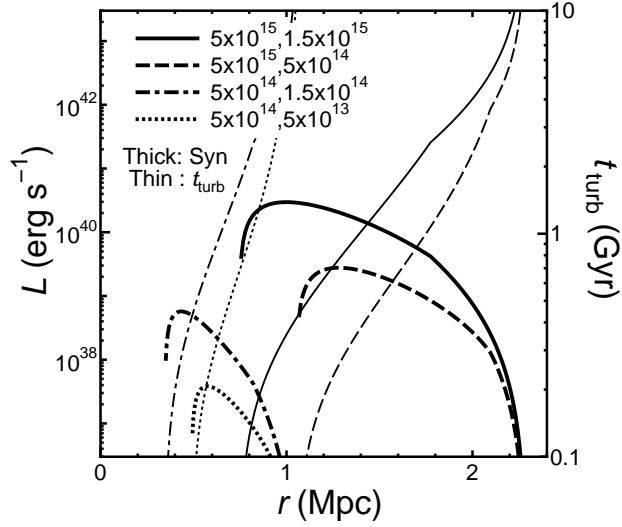} \caption{Thin lines are the time-scale for decay of fluid
 turbulence when the smaller cluster is at the distance $r$ from the
 center of the larger cluster. Thick lines are the synchrotron
 luminosities (same as those in Figure~\ref{fig:Asi}).
 Notation is the same as Figure~\protect\ref{fig:Asi}.
\label{fig:Atturb}}
\end{figure}

\begin{figure}\epsscale{1.00}
\plottwo{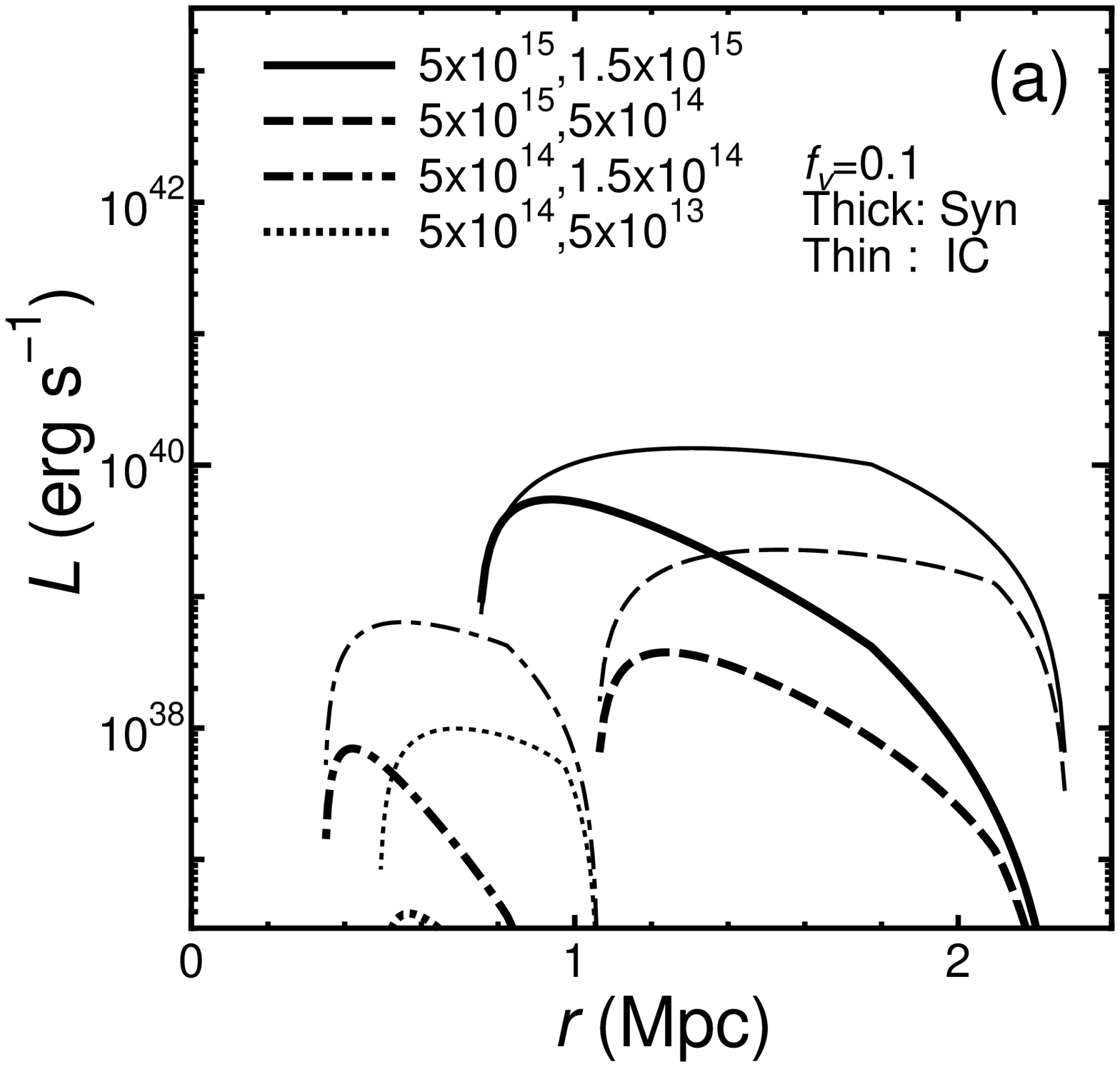}{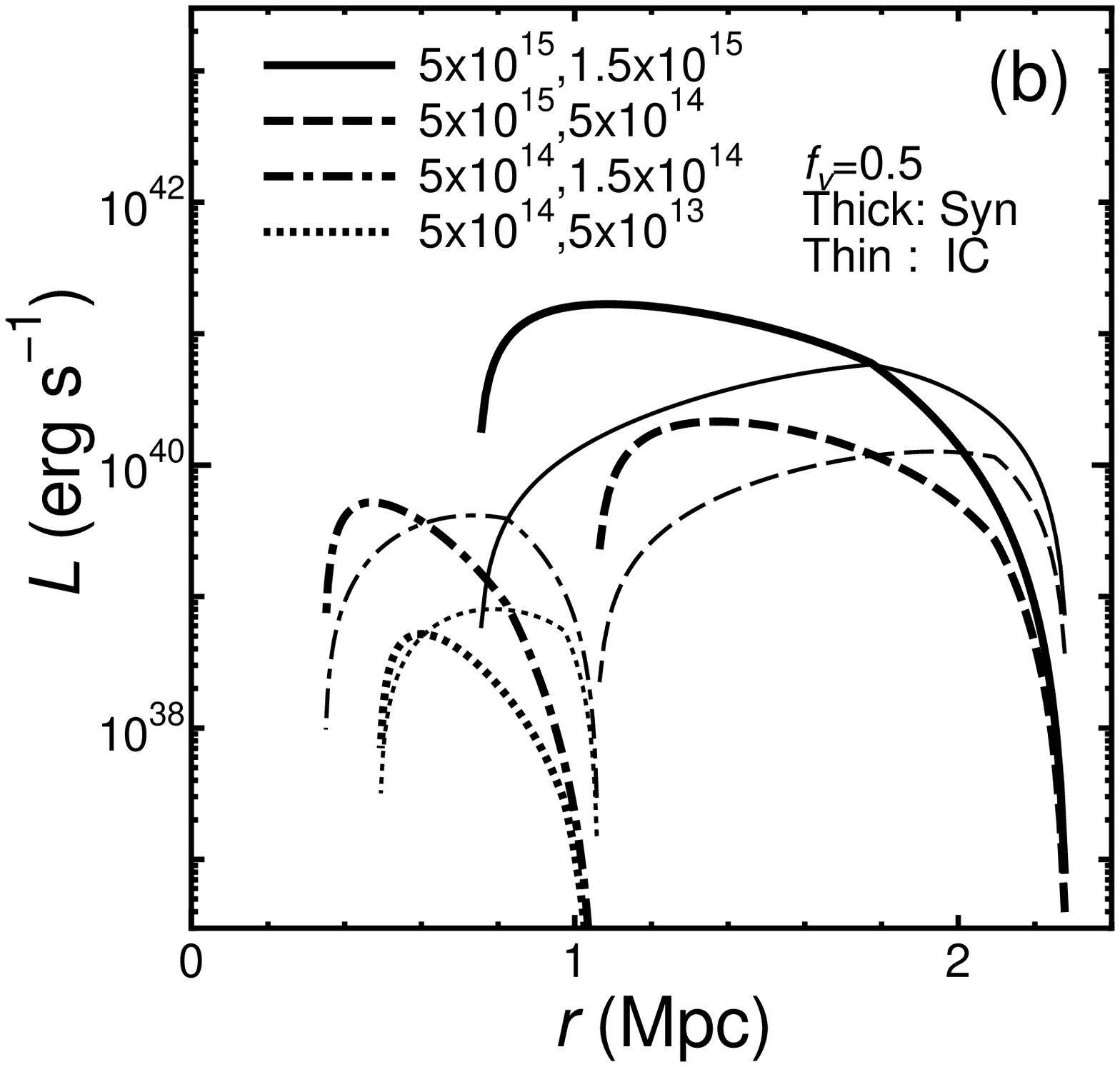} \caption{Same as Figure~\ref{fig:Asi} but
 for (a) Model~B1 (b) Model~B2. \label{fig:Bsi}}
\end{figure}

\begin{figure}\epsscale{0.50}
\plotone{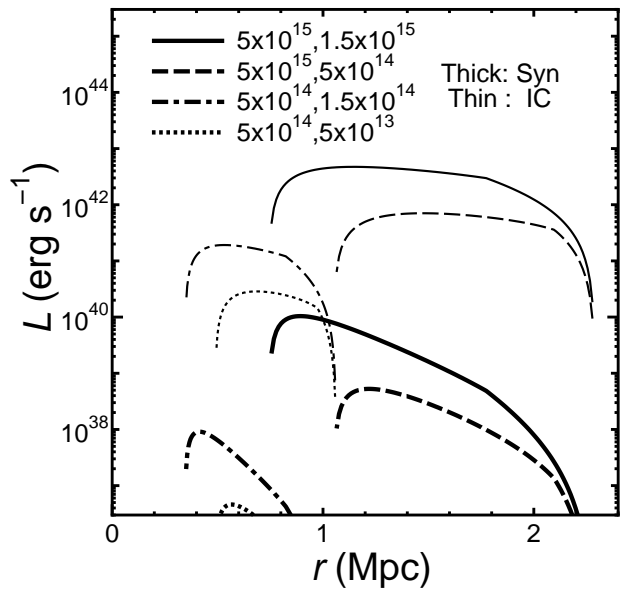} \caption{Same as Figure~\ref{fig:Asi} but
 for Model~C. \label{fig:Csi}}
\end{figure}

The magnetic fields shown in Figure~\ref{fig:Amag} are roughly
consistent with the large values implied by Faraday rotation
measurements \citep[e.g.,][]{law82,kim90,gol93,cla01}. However, the
magnetic fields must be much smaller than the values implied by the
Faraday rotation measurements if both nonthermal radio and hard X-ray
emission are attributed to the same electron population with a power-law
energy distribution without an upper cutoff \citep{fus99,fus00}. Thus,
we also consider the case where the magnetic fields are small.
Figure~\ref{fig:Csi} shows the nonthermal emission when $b=0.001$. In
this case, $B$ is about 30 times smaller than that in
Figure~\ref{fig:Amag}. Although the synchrotron luminosity is not much
different from that in Figure~\ref{fig:Asi}, the inverse Compton
luminosity is much larger. This is because the total nonthermal
luminosity becomes larger (eqs.~[\ref{eq:PA}] and~[\ref{eq:lT}]) and the
fraction of Compton luminosity also becomes larger
(eq.~[\ref{eq:Lcom}]). Since the maximum inverse Compton luminosity is
relatively large ($\sim 10^{42-43}\rm\; erg\; s^{-1}$), the observations
of both nonthermal radio and hard X-ray emission may be explained by
this turbulence resonant acceleration model. If these small magnetic
fields are correct, then the turbulent region must be different from the
region where the Faraday rotation is measured.

\begin{figure}\epsscale{1.00}
\plottwo{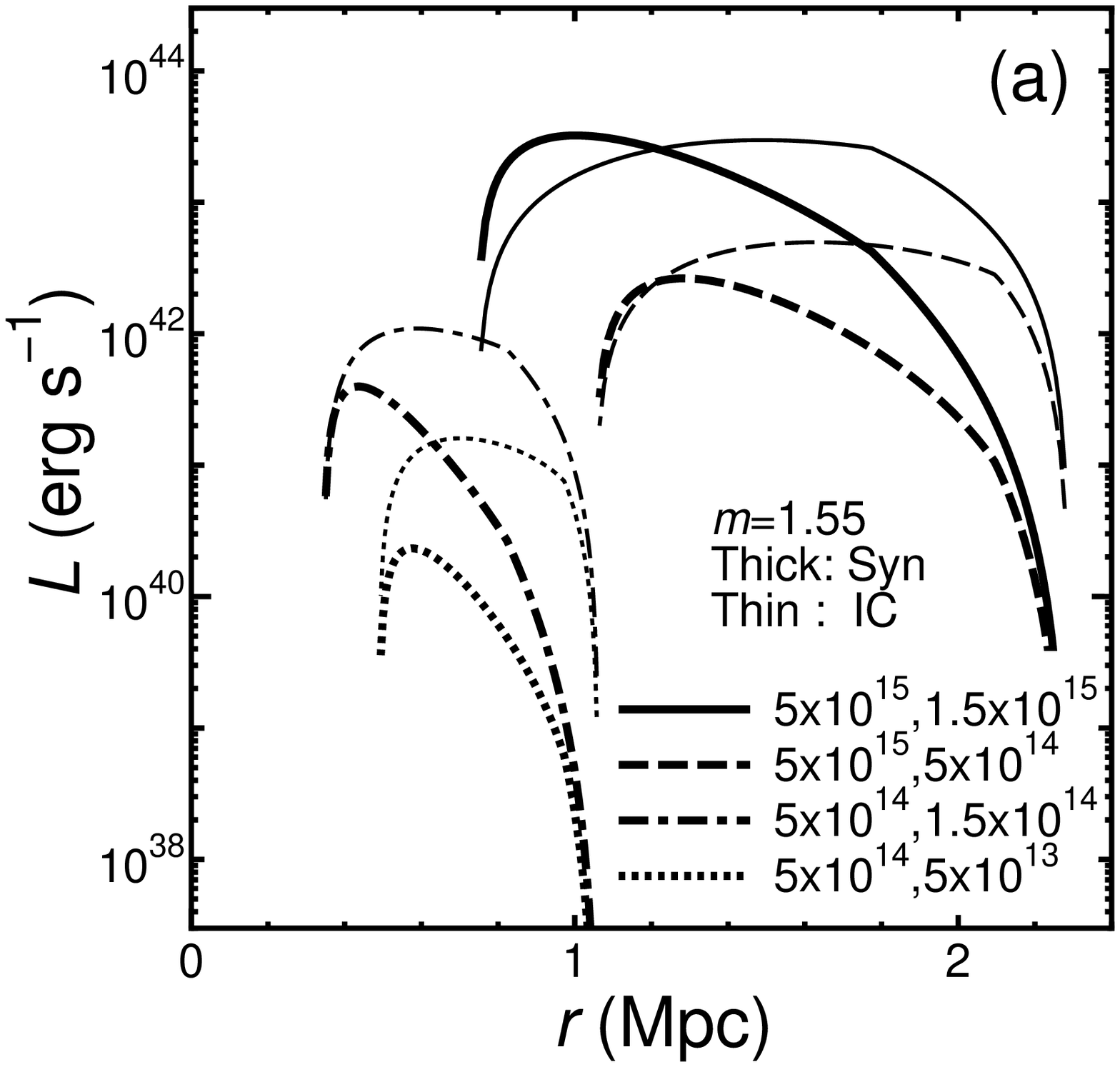}{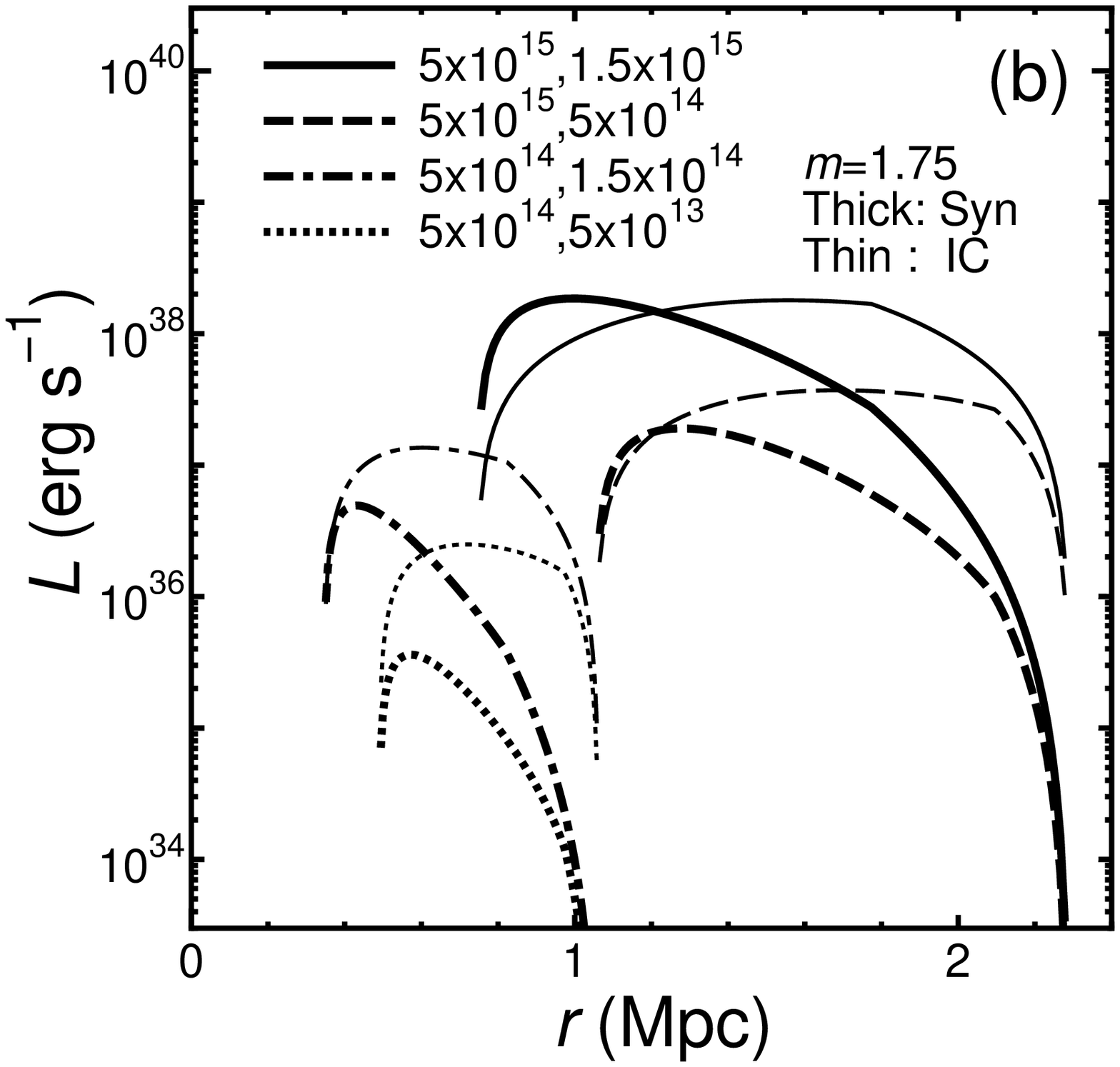} \caption{Same as Figure~\ref{fig:Asi} but
 for (a) Model~D1 (b) Model~D2. \label{fig:Dsi}}
\end{figure}

In \S\ref{sec:gen}, we assumed that the fluid turbulence follows the
Kolmogorov law ($m=5/3$). However, for fully developed MHD turbulence,
\citet{iro63} and~\citet{kra65} suggested that the spectral index should
be $m=3/2$, although this is controversial \citep{gol97}. On the other
hand, if turbulence has not had time to establish a Kolmogorov cascade,
the index may be $m>5/3$. Thus, we discuss the $m$-dependence of the
nonthermal emission. In Figure~\ref{fig:Dsi}, we present the nonthermal
emission when $m=1.55$ and~1.75. It shows that the emission is very
sensitive to $m$. When $m=1.55$, the inverse Compton emission reaches
$\sim 10^{43-44}\;\rm erg\; s^{-1}$, which is consistent with the
observed hard X-ray luminosities \citep{fus99,fus00} even though the
magnetic field is strong ($b = 1 $).  (Note that if we take $m=1.5$, the
inverse Compton luminosity is much larger than the observed
luminosities.) On the other hand, the synchrotron luminosity exceeds the
observed values of $\lesssim 10^{41}\;\rm erg\; s^{-1}$. However, if the
maximum energy of electrons is set by fluid viscosity at $\gamma\sim
10^4$ (eq.~[\ref{eq:gmax}]), the observed synchrotron luminosity at
$\sim 1$~GHz may be smaller than that shown in Figure~\ref{fig:Dsi}
while the inverse Compton luminosity is almost the same
\citep{bru01a,pet01,fuj01d}. Thus, the model with $m<5/3$ can explain
both the nonthermal radio and inverse Compton emission as a result of
particle acceleration by turbulence.

\begin{figure}\epsscale{0.45}
\plotone{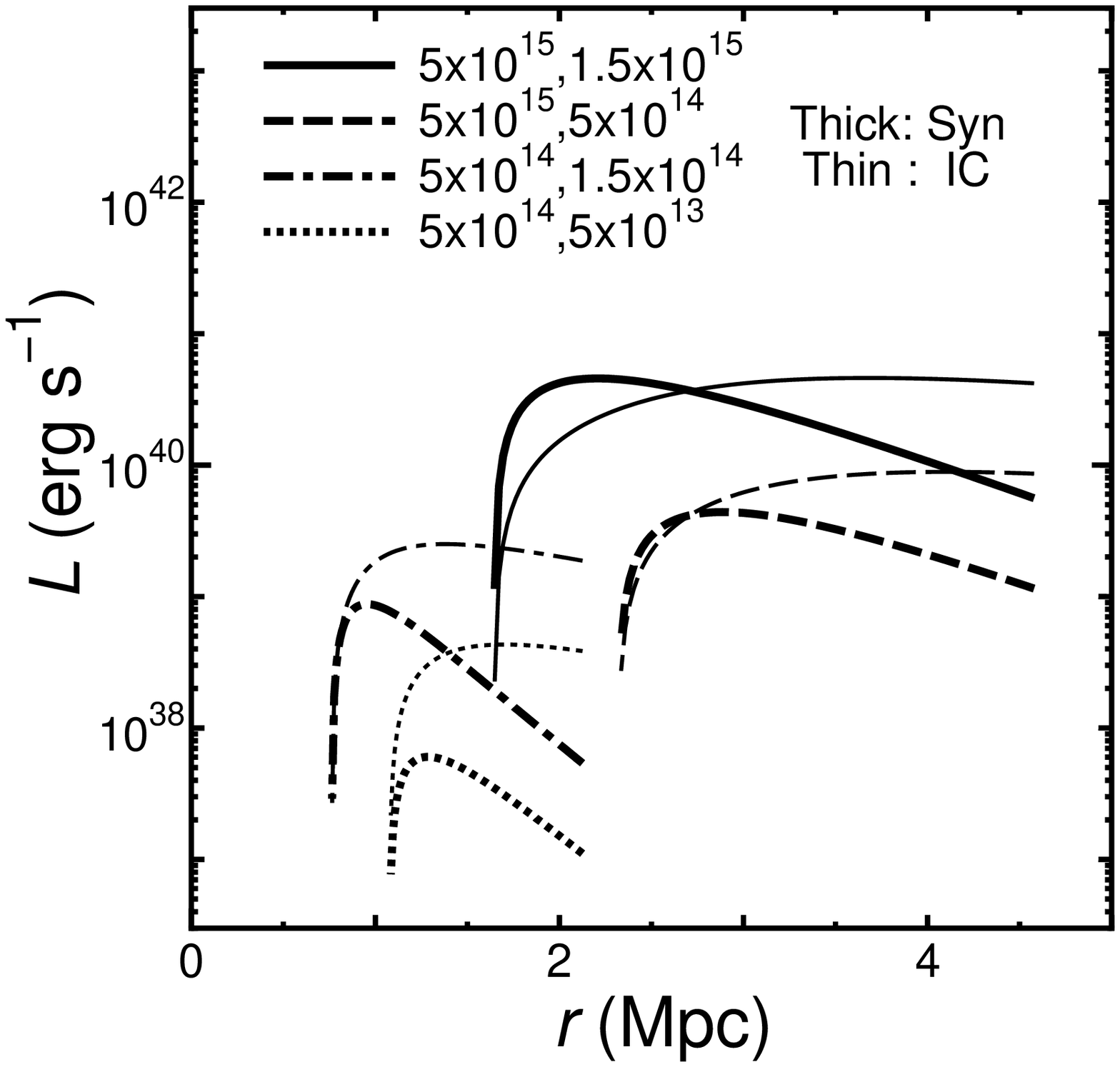} \caption{Same as Figure~\ref{fig:Asi} but
 for Model~E ($x_i=2$). \label{fig:Esi}}
\end{figure}

\begin{figure}\epsscale{0.50}
\plotone{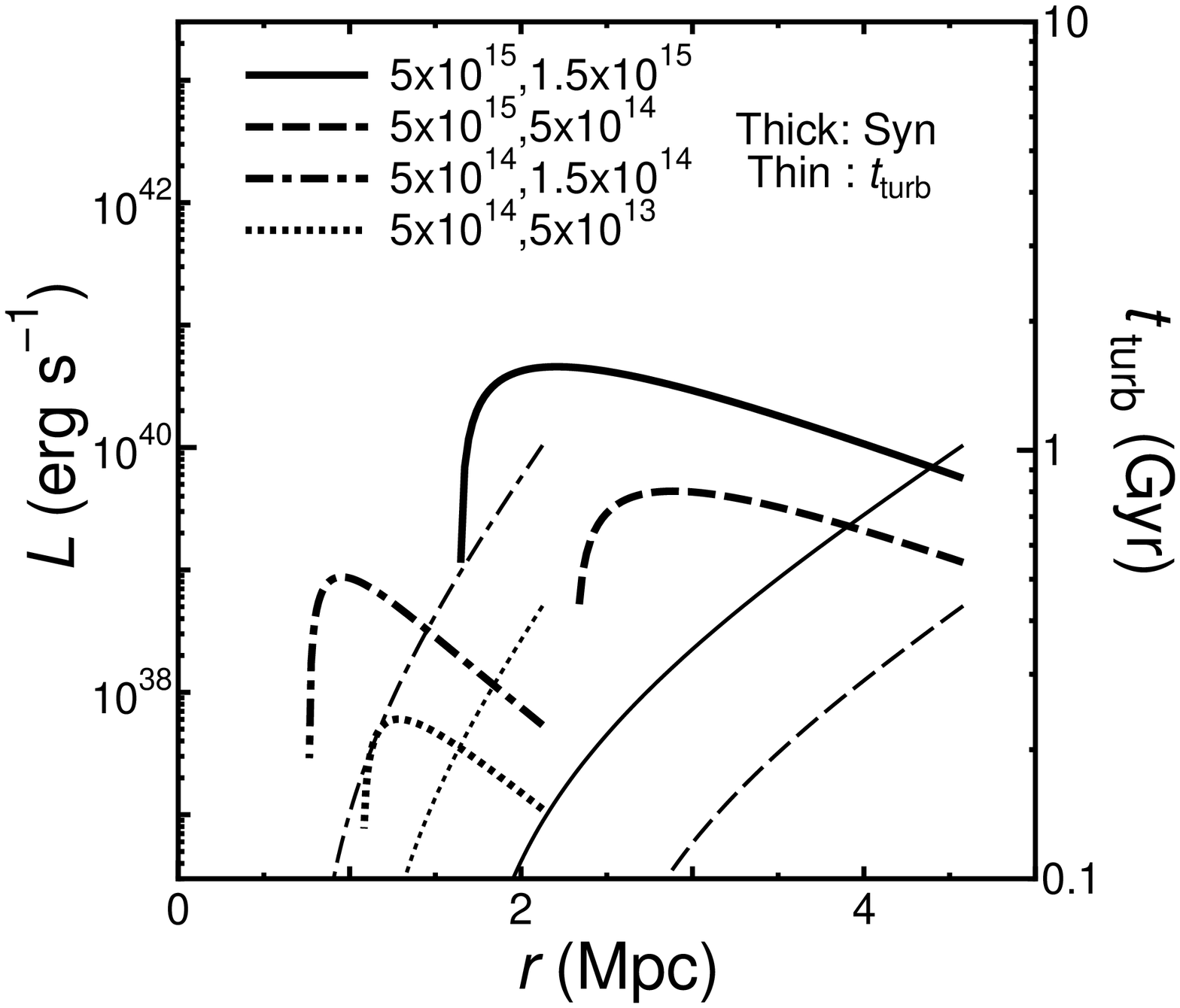} \caption{Same as Figure~\ref{fig:Atturb} but
 for Model~E ($x_i=2$). \label{fig:Etturb}}
\end{figure}

\begin{figure}\epsscale{0.45}
\plotone{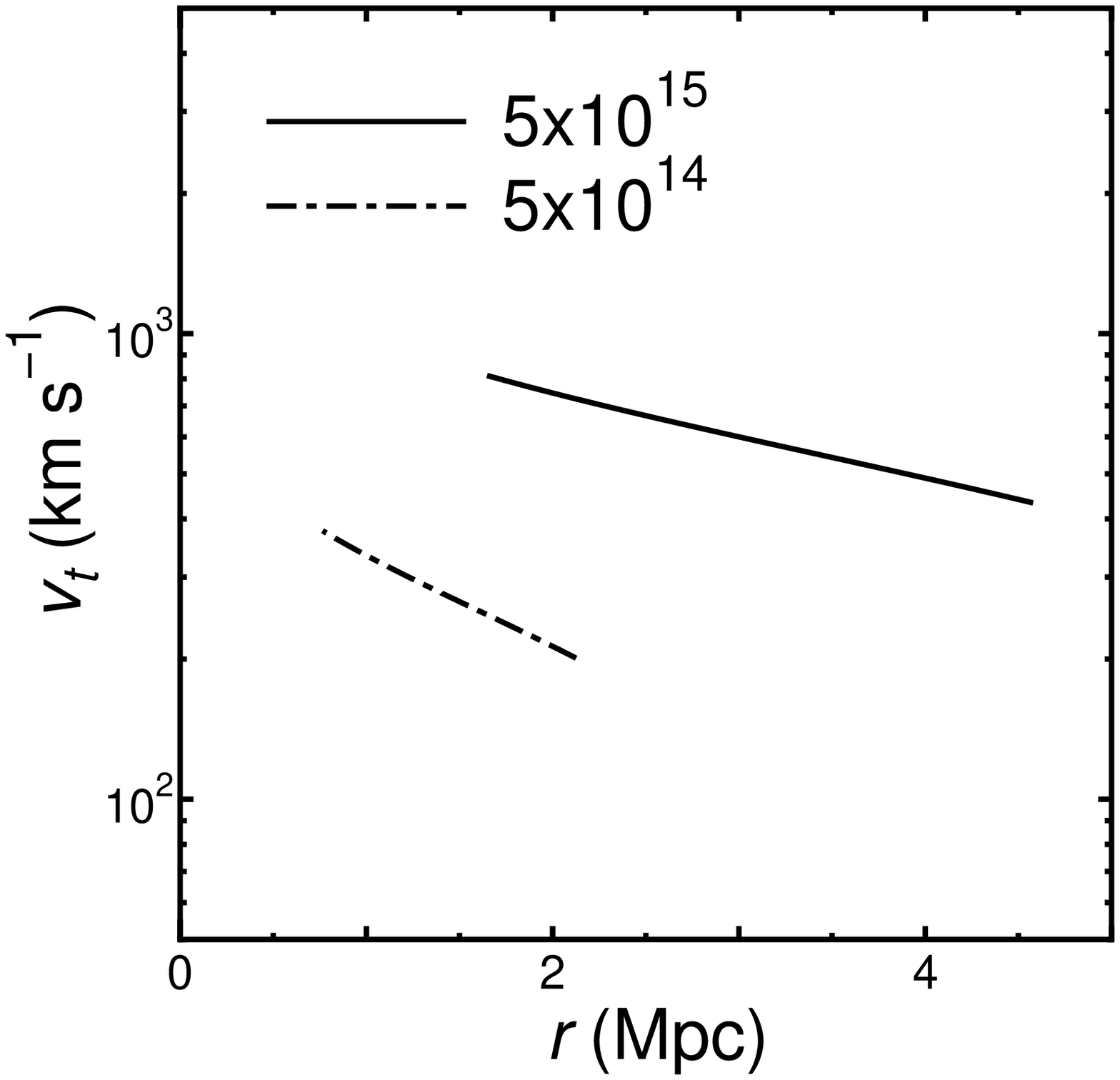} \caption{Same as Figure~\ref{fig:Avt} but
 for Model~E ($x_i=2$). \label{fig:Evt}}
\end{figure}

Although the possibility is small \citep{fuj02b}, a smaller cluster may
fall from the turnaround radius of a larger cluster ($x_i=2$). In this
case, the smaller cluster has a large velocity near the center of the
larger cluster compared to the case in
\S\ref{sec:gen}. Figure~\ref{fig:Esi} shows the nonthermal emission from
the merging clusters. After the smaller cluster enters the virial radius
of the larger cluster, it produces nonthermal emission.  The maximum
luminosities are not much different from those in Figure~\ref{fig:Asi}.
Although the smaller cluster has a larger velocity, the radius is
smaller owing to the larger ram-pressure.  However, as shown in
Figure~\ref{fig:Etturb}, the time-scale of the fluid turbulence is small
($t_{\rm turb}\lesssim 10^9$~yr). This means that the turbulence cannot
accelerate electrons at $\gamma\sim 10^3$ up to $\sim 10^4$, where the
observed nonthermal emission is produced. Thus, the actual luminosities
may even smaller than those shown in Figure~\ref{fig:Esi}. On the other
hand, since the turbulent velocity is large ($\sim 10^3\rm\; km\;
s^{-1}$; Figure~\ref{fig:Evt}), it would be easily observable by high
spectral resolution X-ray detectors (e.g., {\it Astro-E2}).

\subsection{Preheating}
\label{sec:preh}

If the gas distribution of the small cluster has been flattened and the
central gas density had been decreased by non-gravitational heating
before cluster formation (preheating), the gas is more easily removed by
the ram-pressure when the smaller cluster plunges into the larger
cluster.  Figure~\ref{fig:Esi} shows how the nonthermal emission from
merging clusters is affected by preheating.  Compared with the case of
no preheating (Figure~\ref{fig:Asi}), the maximum luminosities of the
nonthermal emission is small especially when the mass of the smaller
cluster is small. This is because preheating affects less massive
clusters significantly and the gas is stripped earlier especially when
the virial temperature of the smaller of the merging clusters approaches
$T_1$ (eqs.~[\ref{eq:TT}] and~[\ref{eq:ICM_H}]).

Radio observations have shown that the luminosity of diffuse radio
emission from clusters is a steep function of cluster X-ray temperature
or luminosity \citep{fer00,lia00}. Since the sample of radio emission is
small and less luminous radio clusters may not have been observed, this
correlation may actually show that the maximum radio luminosity for a
given X-ray temperature or luminosity is a steep function of cluster
X-ray temperature or luminosity. This suggests that the steep
correlation may apply only to clusters undergoing a major merger (small
$M_{L, \rm vir}/M_{S, \rm vir}$).

Since the preheating reduces the nonthermal emission from less massive
clusters, it may help to explain the observed steep correlation.  We
assume that $\min(M_{L, \rm vir}/M_{S, \rm vir})\sim 3$ and it does not
depend on $M_{L, \rm vir}$. For the no preheating model
(Figure~\ref{fig:Asi}), the synchrotron luminosity from merging clusters
of $M_{L, \rm vir}=5\times 10^{14} M_{\sun}$ and $M_{S, \rm
vir}=1.5\times 10^{14} M_{\sun}$ is about 50 times smaller than that
from merging clusters of $M_{L, \rm vir}=5\times 10^{15} M_{\sun}$ and
$M_{S, \rm vir}=1.5\times 10^{15} M_{\sun}$. If the temperature of the
larger clusters are represented by $T_{\rm ICM}\propto M_{L, \rm
vir}^{2/3}$ \citep*[e.g.,][]{evr96}, the ratio of the temperature
between the two cases is 4.6. Thus, the synchrotron luminosity from
merging clusters roughly follows $L_{\rm syn}\propto T_{\rm ICM}^{2.6}$.
Similarly, for the preheating model (Figure~\ref{fig:Esi}) we obtain
$L_{\rm syn}\propto T_{\rm ICM}^{3.0}$. Since the virial temperature of
the smaller cluster of $M_{S, \rm vir}=1.5\times 10^{14} M_{\sun}$ is
$\sim T_1$, its merger with the larger cluster is significantly affected
by the preheating. However, although preheating steepens the relation,
the power-law index is still smaller than that observed \citep[$\sim
6$;][]{fer00}.

\subsection{Turbulence Developed in the Whole Cluster}
\label{sec:whole}

Numerical simulation done by \citet{roe99} showed that the fluid
turbulence in ICM is also developed after a smaller cluster passes the
center of a larger cluster. This turbulence is pumped by dark
matter-driven oscillations in the gravitational potential; the
merger-induced large scale bulk flows breakdown into turbulent gas
motions. In this case, we expect the turbulence prevails on a cluster
scale, and we can roughly estimate the nonthermal emission from the
cluster. If we assume that $v_t=1000\rm\; km\: s^{-1}$, $l_0=1$~Mpc,
$\rho_{\rm ICM}=7 \times 10^{-28}\rm\: g\: cm^{-3}$, $T_{\rm
ICM}=10^8$~K, $b=1$, and $m=5/3$, we obtain $L_{\rm syn}=8\times
10^{41}\rm\: erg\: s^{-1}$ and $L_{\rm IC}=1\times 10^{41}\rm\: erg\:
s^{-1}$ using the model in \S\ref{sec:model}.  Thus, turbulence is
strong enough to produce the observed synchrotron emission ($\sim
10^{41}\rm\: erg\: s^{-1}$), but cannot produce the observed hard X-ray
emission ($\sim 10^{43-44}\rm\: erg\: s^{-1}$) as long as $m=5/3$ and
the magnetic fields are strong ($b = 1$). However, for the same
parameters but with weak magnetic fields ($b=0.001$), we obtain $L_{\rm
syn}=8\times 10^{41}\rm\: erg\: s^{-1}$ and $L_{\rm IC}=9\times
10^{43}\rm\: erg\: s^{-1}$. The luminosities are comparable to the
observations. Observed central radio halos may be this kind of
nonthermal emission from the whole cluster. The turbulent velocity,
$v_t$, would be large enough to be detected by {\it Astro-E2}.

\section{Conclusions}
\label{sec:conc}

We have investigated nonthermal emission from electrons accelerated by
resonant Alfv\'{e}n waves in clusters of galaxies. We assume that the
Alfv\'{e}n waves are driven by fluid turbulence generated by cluster
mergers.  We find that the resonant Alfv\'{e}n waves can accelerate
electrons up to $\gamma\sim 10^5$; this value is limited by fluid
viscosity. Our calculations show that the turbulent resonant
acceleration can give enough energy to electrons to produce the observed
diffuse radio emission from clusters if there is a pool of electrons of
$\gamma\sim 10^3$ in clusters.. On the other hand, the observed hard
X-ray emission from clusters is explained by the turbulent resonant
acceleration only when magnetic fields are small ($\lesssim \mu G$) or
the fluid turbulence spectrum is flatter than the Kolmogorov law. The
fluid turbulence responsible for the particle acceleration would be
observed by {\it Astro-E2} in the regions where diffuse radio emission
is observed. Although non-gravitational heating before cluster formation
(preheating) makes the relation between radio luminosity and X-ray
temperature steeper, our predicted relation is still flatter than the
observed one.

\acknowledgments

We thank H. Ohno, H. Matsumoto, and T.~N. Kato for useful discussions.
Y.~F.\ and M.~T.\ were each supported in part by a Grant-in-Aid from the
Ministry of Education, Science, Sports, and Culture of Japan (Y.~F.:
14740175; M.~T.: 13440061). C. L. S. was supported in part by $Chandra$
Award Number GO1-2123X, issued by the $Chandra$ X-ray Observatory
Center, which is operated by the Smithsonian Astrophysical Observatory
for and on behalf of NASA under contract NAS8-39073, and by NASA XMM
Grant NAG5-10075.

\clearpage

\begin{figure}\epsscale{0.45}
\plotone{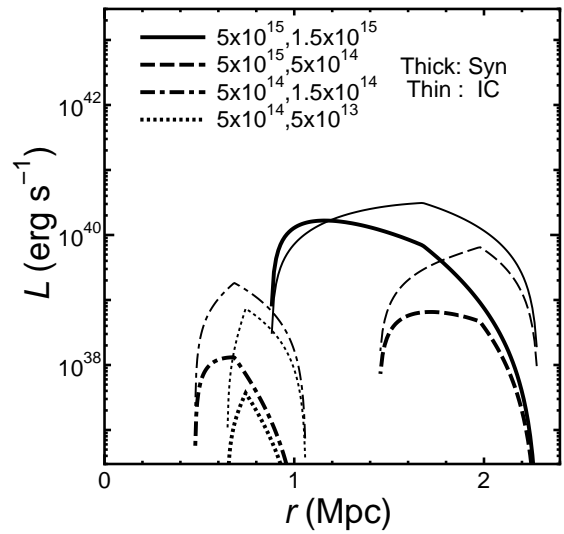} \caption{Same as Figure~\ref{fig:Asi} but
 for Model~F (preheating). \label{fig:Fsi}}
\end{figure}


\begin{deluxetable}{cccccc}
\footnotesize
\tablecaption{Model Parameters\label{tab:para}}
\tablewidth{0pt}
\tablehead{
\colhead{Models} 
   & \colhead{$f_v$}
         & \colhead{$b$}  
               & \colhead{$m$} 
                     & \colhead{$x_{\rm i}$} 
                           & \colhead{Preheating}  
}
\startdata
A  & 0.2 & 1   & 5/3 & 0.5 & no \\
B1 & 0.1 & 1   & 5/3 & 0.5 & no \\
B2 & 0.5 & 1   & 5/3 & 0.5 & no \\
C  & 0.2 &0.001& 5/3 & 0.5 & no \\
D1 & 0.2 & 1   & 1.55& 0.5 & no \\
D2 & 0.2 & 1   & 1.75& 0.5 & no \\
E  & 0.2 & 1   & 5/3 &  2  & no \\
F  & 0.2 & 1   & 5/3 & 0.5 & yes \\
\enddata
\end{deluxetable}

\end{document}